\documentclass[12pt]{article}
\newcommand\BibTeX{{\rmfamily B\kern-.05em \textsc{i\kern-.025em b}\kern-.08em
T\kern-.1667em\lower.7ex\hbox{E}\kern-.125emX}}
\usepackage{graphicx,verbatim,array,multicol, subfigure, color, lscape, mathrsfs}
\usepackage{psfrag, amsmath, amsfonts, epsfig, fancybox, setspace,soul, lscape,bm,hyperref}
\usepackage{xr}

\usepackage[english]{babel}
\usepackage[square,numbers]{natbib}

\def\supg{^{(g)}}

\def\bbeta{\boldsymbol{\beta}}

     \def\bW{\mathbf{W}}

          \def\bB{\mathbf{B}}
\def\bw{\mathbf{w}}

\def\Dscr{\mathscr{D}}

\def\supone{^{(1)}}
\def\supzero{^{(0)}}
\def\suponep{^{(1p)}}
\def\supzerop{^{(0p)}}

\definecolor{green}{RGB}{000,150,100}
\definecolor{purple}{RGB}{150,000,180}

\begin{document}
\thispagestyle{empty}
\begin{center}
\vspace*{25mm}
\LARGE Using a Surrogate with Heterogeneous Utility to Test for a Treatment Effect\\
\vspace*{15mm}
\normalsize Layla Parast$^{1}$, Tianxi Cai$^{2}$, and Lu Tian$^{3}$\\
\vspace*{10mm}
\normalsize $^{1}$Statistics Group, RAND Corporation, 1776 Main Street,  Santa Monica, CA 90401\\
\normalsize $^{2}$Department of Biostatistics, Harvard University, 665 Huntington Avenue, Boston, Massachusetts 02115\\
\normalsize $^{3}$Department of Biomedical Data Science, Stanford University,  Stanford, CA 94305\\
\end{center}

\clearpage
\begin{abstract}
The primary benefit of identifying a valid surrogate marker is the ability to use it in a future trial to test for a treatment effect with shorter follow-up time or less cost. However, previous work has demonstrated potential heterogeneity in the utility of a surrogate marker. When such heterogeneity exists, existing methods that use the surrogate to test for a treatment effect while ignoring this heterogeneity may lead to inaccurate conclusions about the treatment effect, particularly when the patient population in the new study has a different mix of characteristics than the study used to evaluate the utility of the surrogate marker. In this paper, we develop a novel test for a treatment effect using surrogate marker information that accounts for heterogeneity in the utility of the surrogate. We compare our testing procedure to a test that uses primary outcome information (gold standard) and a test that uses surrogate marker information, but ignores heterogeneity. We demonstrate the validity of our approach and derive the asymptotic properties of our estimator and variance estimates. Simulation studies examine the finite sample properties of our testing procedure and demonstrate when our proposed approach can outperform the testing approach that ignores heterogeneity. We illustrate our methods using data from an AIDS clinical trial to test for a treatment effect using CD4 count as a surrogate marker for RNA.

\thispagestyle{empty}
\end{abstract}

\clearpage

\section{Introduction}

There has been a substantial growth in clinical and methodological research on identifying and using valid surrogate markers in the past few decades. A valid surrogate marker is a biological measurement that can be used as a replacement for a primary outcome of interest in a clinical study. Many statistical methods have been proposed to evaluate and validate surrogate markers using a wide variety of innovative methodological approaches.\cite{prentice1989surrogate,burzykowski2005evaluation,wang2002measure,gilbert2008evaluating,parast2015robust} The primary benefit of identifying a valid surrogate marker is the ability to use it in a future trial to test for a treatment effect with less required follow-up time or less cost. For example, the U.S. Food and Drug Administration announced in 2020 that a surrogate marker that could be measured earlier than COVID-19 infection could be used to assess the vaccine efficacy in preventing infection,\cite{avorn2020up} thus potentially allowing for earlier identification of effective vaccines.

Several statistical methods have been proposed in recent years to assess the treatment effect on the primary outcome based on surrogate marker information. For example, Parast et al. (2019)\cite{parast2019using} proposed a nonparametric approach to test for a treatment effect in a time-to-event outcome setting based on a surrogate marker measured at an earlier time point utilizing information about the relationship between the surrogate marker and primary outcome obtained from a prior study. Chen et al. (2020)\cite{chen2020model} suggested a model-based approach that uses surrogate information to make interim decisions about whether to drop a treatment arm or stop a trial for futility. Price et al. (2018)\cite{price2018estimation} defined an optimal surrogate that optimally predicts a primary outcome and proposed super-learner and targeted super-learner based estimation procedures. Athey et al. (2019)\cite{athey2019surrogate} proposed to combine multiple surrogate markers to predict a long term outcome and estimate a treatment effect, and explicitly characterized the difference between the treatment effect estimated based on the primary outcome versus the surrogate combination.

Previous clinical and methodological work has demonstrated potential heterogeneity in the utility of a surrogate marker i.e. that a surrogate marker may be more useful (with respect to capturing the treatment effect on the primary outcome) for some subgroups than for others.\cite{lin1993evaluating} Parast et al. (2021)\cite{parast2021het} offers a nonparametric estimation procedure and formal test for heterogeneity of surrogate utility with respect to a baseline covariate. When such heterogeneity exists, existing methods that use the surrogate to test for a treatment effect while ignoring this heterogeneity may lead to inaccurate conclusions about the treatment effect, particularly when the patient population in the current study has a different mix of characteristics than the prior study (used to evaluate the utility of the surrogate marker).

For example, in the simulation study in this paper, we examine a setting where the estimated treatment effect based on the primary outcome is 33.7 (standard error [SE] = 1.6); applying the testing approach of Parast et al. (2019)\cite{parast2019using} which uses surrogate marker information but does not account for heterogeneity, the estimated treatment effect on the primary outcome is 39.2 (SE=3.5). The approach of Parast et al. (2019)\cite{parast2019using} guarantees that the treatment effect based on the surrogate will be a lower bound for the true treatment effect on the primary outcome under certain conditions. However, these conditions may be violated when there is heterogeneity in the utility of the surrogate and thus leads to this type of situation where the estimated treatment effect using the surrogate is much higher than that using the primary outcome. Our approach that we propose in this paper which incorporates heterogeneity produces a treatment effect estimate that retains the lower bound property, with similar power to the treatment effect using the primary outcome. While we focus on heterogeneity with respect to a continuous baseline covariate, we provide a motivational example in Appendix A where there is heterogeneity with respect to a discrete covariate, gender. In this example, the surrogate marker is strong among males (explaining 99\% of the treatment effect on the primary outcome) but weaker among females (explaining 67\%). In a new study where the distribution of gender is 95\% female and 5\% male and the treatment effect on the primary outcome is 38.95, using the surrogate marker and accounting for heterogeneity in surrogacy produces an estimated treatment effect on the primary outcome equal to 17.95 while ignoring heterogeneity produces an estimate of 44.5, again, failing to correctly provide a lower bound on the true treatment effect. In contrast, if we consider a future study where the distribution of gender is 5\% female and 95\% male, the treatment effect on the primary outcome is 74.05, while the treatment effect using the surrogate and accounting for heterogeneity is 71.05 versus not accounting for heterogeneity is 44.5, indicating a potential loss in power to detect a treatment effect when heterogeneity is ignored.

In this paper, we develop a novel test for a treatment effect using surrogate marker information that accounts for heterogeneity in the utility of the surrogate. We compare our testing procedure to a test that uses primary outcome information only (gold standard) and a test that uses surrogate marker information, but ignores heterogeneity. We demonstrate the validity of our testing procedure and derive the asymptotic properties of our estimator and variance estimates. A simulation study is used to examine the finite sample properties of our testing procedure and demonstrate when our proposed approach can outperform the testing approach that ignores heterogeneity. {In particular, we demonstrate examples where the test of Parast et al. (2019)\cite{parast2019using} provides an incorrect estimate with respect to the treatment effect.} We illustrate our approach using data from an AIDS clinical trial to test for a treatment effect using CD4 count as a surrogate marker for plasma HIV-1 RNA.

\section{Testing Procedure}
\subsection{Notation and Setting}
We focus on a setting where we are currently conducting a study to examine the effect of a treatment on a primary outcome of interest, denoted by $Y$, and we additionally have data available from a {prior} study. We assume that this prior study was used to examine the strength of the surrogate, denoted by $S$, and heterogeneity in the utility of the surrogate, and has measurements of both $Y$ and $S$ of the current study. Let $Z$ denote the treatment indicators where treatment is randomized and $Z \in \{0,1\}$ (i.e., treatment vs. control), and $W$ denote a baseline covariate such that $S$ has been shown to have heterogeneous utility with respect to this covariate. Without loss of generality, we take $W$ to be continuous; all proposed procedures can easily accommodate a discrete $W$ as well. {{We focus on a setting with heterogeneity with respect to a single baseline covariate $W$; in Section \ref{multiple}, we discuss an extension to multiple $W$.}} In addition, we assume we are in a setting where either $S$ is measured earlier than $Y$ or $S$ is measured at the same time as $Y$ but is less expensive, invasive or burdensome, and there is no censoring or missing data. {{Throughout this paper, we quantify surrogate strength/utility using the quantity: the proportion of treatment effect on the primary outcome explained by the treatment effect on the surrogate marker. \cite{freedman1992statistical,wang2002measure,parast2015robust}}} We use potential outcomes notation where each person has a potential $\{Y \supone, Y \supzero, S \supone, S \supzero\}$ where $Y \supg$ is the outcome when $Z=g$ and $S \supg$ is the surrogate when $Z=g$. Observed data from the current study is denoted as and consists of $\Dscr=\{ (Y_{gi}, S_{gi}, W_{gi}), i=1,...,n_g; g=0,1\}$, where $n_g$ denotes the number of individuals in treatment group $g$.

The goal in the current study is to test for a treatment effect on the primary outcome quantified as $$H_0: \Delta \equiv E(Y \supone - Y \supzero) = E(Y \supone) - E(Y \supzero) = 0.$$

\noindent Our aim is to leverage information from the \textit{prior} study to test $H_0$ using surrogate marker information in order to reduce study follow-up time, costs, and/or participant burden, i.e., making inference on $\Delta$ without using $\{Y_{gi}, i=1,\cdots, 1, n_g; g=0, 1\}.$ We use a superscript $p$ to denote ``prior" when referring to data or quantities from the prior study. For example, we denote observed data from the prior study by $\Dscr^p=\{ (Y^p_{gi}, S^p_{gi}, W^p_{gi}, i=1,...,n_g^p, g=0,1\}$, where $n_g^p$ is the sample size of treatment group $g$.

\subsection{Assumptions} \label{assumptions}
Given that our setting rests on the existence of a valid surrogate marker, we first define $S$ to be a valid surrogate marker for $Y$ if the following conditions hold:
\begin{enumerate}
\item[] (C1) $E(Y^{(0)}  | S^{(0)}=s, W=w)$ is a monotone function of $s;$
\item[] (C2) $P(S^{(1)} >s | W=w) \ge P(S^{(0)}  >s | W=w)$ for all $s$ and $w;$
\item[] (C3) $E(Y^{(1)} | S^{(1)}=s,W=w) \ge E(Y^{(0)} | S^{(0)}=s,W=w) $ for all $s$ and $w.$
\item[] (C4) A large proportion of the treatment effect on the primary outcome can be explained by the treatment effect on the surrogate marker for all $w$.
\end{enumerate}
Assumptions (C1)-(C3) are parallel to those required in Wang and Taylor (2002)\cite{wang2002measure} and Parast et al. (2017)\cite{parast2016nonparametric} and protect against the surrogate paradox situation. \cite{vanderweele2013surrogate}  Assumption (C1) implies that the surrogate marker is either ``positively" or ``negatively" related to the time of the primary outcome, (C2) implies that there is a positive treatment effect on the surrogate marker, and (C3) implies that there is a non-negative residual treatment effect beyond that on the surrogate marker. Assumptions (C1-C3) together guarantee that $E(Y\supone \mid W=w)\ge E( Y \supzero \mid W=w)$, for all $w$ in the support of $W$ (see Appendix B). {{Lastly, (C4) states that the proportion of the treatment effect explained by the surrogate marker must be large and guarantees the strength of the surrogate marker of interest for all individuals in the study. While this is somewhat vague, there is no agreed upon value that signifies a ``large” proportion, though previous work has tended to view values of 0.6-0.75 or higher as large. \cite{lin1997estimating,freedman1992statistical,eastell2003relationship}} }If the existing heterogeneity is such that the surrogate is strong for some $w$ and weak for other $w$, it should \textit{not} be used as a replacement of the primary outcome for all individuals in a future study. Instead, one may consider using the surrogate as a replacement only among those with a $W$ where the surrogate is strong; we discuss this further in the Discussion.

In order to ensure that the proposed test statistic to be described in Section \ref{test-statistic}, has a reasonable interpretation with respect to $\Delta$,  we also require:
\begin{enumerate}
\item[] (C5) $E(Y^{(0)} | S^{(0)}=s, W=w) = E(Y^{(0p)} | S^{(0p)}=s, W^p=w)  $ for all $s$ and $w$;
\item[] (C6) $E(Y^{(0p)} | S^{(0p)}=s, W^p=w)$ is estimable for any $(s, w)\in \Omega_J,$  where $\Omega_J$ is the common compact support for both $(S^{(g)},W^{(g)})$ in $g=0,1.$
\end{enumerate}
\def\Hdag{H^{\dag}}
\noindent Assumption (C5) implies that in the control groups, the current study and the prior study share the same conditional expectation for $Y$ given $S$ and $W$.  This assumption is reasonable when, for example, the control condition in both studies are  the same, such as ``usual care." Importantly, such an assumption is not required to hold for the treatment groups and it relaxes the requirement that the distribution of $Y$ conditional on $S$ be transportable from the prior to current study. {{Even so, this assumption is admittedly very strong and needs to be carefully considered before using this approach; however, \textit{any} testing procedure that attempts to borrow information from a prior study to test a hypothesis in a future study is going to require some type of strong transportability assumption. If there is reason to believe that such transportability between studies is not appropriate, then the prior study should not be considered for informing the future study.}} Assumption (C6) ensures that we can approximate $E(Y^{0}|S^0=s, W^0=w)$ for all observed pairs of $S^{(g)}$ and $W^{(g)}, g=0, 1$ in the current study. We discuss robustness to these assumptions as well as additional assumptions needed for a causal interpretation in Appendix B.

\subsection{Proposed Testing Procedure \label{test-statistic}}
Recall that our aim is to take advantage of information from the prior study to test $H_0$ using surrogate marker information such that this test accounts for known heterogeneity in the utility of the surrogate marker. To achieve this goal we note that $\Delta$ can be expressed as:
\begin{eqnarray}
\Delta &=& E(Y \supone) - E(Y \supzero) = \int  \Delta(w) dF_{W}(w) \nonumber \\
&=& \int \left [\int \mu_1(s,w)dF \supone(s|w)\right ] dF_{W}(w) - \int \left [\int \mu_0(s,w) dF \supzero (s|w)\right ] dF_{W}(w) \label{condexp}
\end{eqnarray}
where  
$\mu_g(s,w) \equiv E(Y \supg | S \supg = s, W = w)$, $F\supg (s|w) \equiv F_{S\supg | W }(s|w)$ is the conditional cumulative distribution function of $S^{(g)}$ given $W=w,$ and $F_W(w)$ is the cumulative distribution of $W. $ In expressing $\Delta$ as (\ref{condexp}), we have simply used a conditional expectation to incorporate $S$ and $W$ into our expression. By expressing $\Delta$ in this way, this motivates the following \textit{earlier} treatment effect definition:
\begin{eqnarray}
\Delta_H &=&  \int \left [\int \mu_0(s,w)dF\supone (s|w)\right ] dF_{W}(w) - \int \left [\int \mu_0(s,w) dF \supzero (s|w)\right ] dF_{W}(w)  \label{deltahdef} \\
&=&  \int \mu_0^p(s,w)dF^{(1)} (s, w) - \int \mu_0^p(s,w) dF^{(0)}(s, w) \label{deltahdef2}
\end{eqnarray}
where $F^{(g)}(s, w)$ is the cumulative distribution function of $(S^{(g)}, W)$ in the current study. The only change in going from (\ref{condexp}) to (\ref{deltahdef}), is that we have replaced $\mu_1(s,w)$ with $\mu_0(s,w)$ in the first term which will ensure that this quantity provides a lower bound on the treatment effect. In the second equality, (\ref{deltahdef2}), we replace  $\mu_0(s,w)$ with $\mu_0^p(s,w)$ which follows from Assumption (C5). The expression  (\ref{deltahdef2}) is now a quantity that only involves $\mu_0^p(s,w)$ which is the conditional risk in the prior study, and the distribution of $S$ and $W$ in the current study. Importantly, the expression does not involve $Y$ from the current study at all. In practice, $\mu_0^p(s, w)$ is unknown and must be replaced with an estimate, $\widehat{\mu}_0^p(s, w)$, which we describe in Section \ref{sub-estimation-delta}. Because of this, we define the following \textit{earlier} average treatment effect quantity, where the \ $\ \widetilde{} \ $ \ notation makes the dependence on information from the prior study explicit:
\begin{eqnarray*}
\widetilde{\Delta}_H &=&  \int \widehat{\mu}_0^p(s,w)dF^{(1)} (s, w) - \int \widehat{\mu}_0^p(s,w) dF^{(0)}(s, w) = E\left\{\widehat{\mu}_0^p(S^{(1)}, W)-\widehat{\mu}_0^p(S^{(0)}, W) \mid \Dscr^p\right\}.
\end{eqnarray*}
This quantity, $\widetilde{\Delta}_H$, measures the treatment effect on a transformation of the surrogate marker and baseline covariate, i.e., the difference between $\widehat{\mu}_0^p(S^{(1)},W)$ and $\widehat{\mu}_0^p(S^{(0)},W).$  First, due to randomization, $W$ has the same distribution between two treatment groups and $\widetilde{\Delta}_H$ has an appealing causal interpretation reflecting the treatment effect on the surrogate marker. Second,  $\widetilde{\Delta}_H$ represents the part of the treatment effect on the primary outcome explained by the surrogate marker and an approximation to $\Delta_H$, which is the quantity of our primary interest. Under the null hypothesis of no average treatment effect on the primary outcome,  there will also be no average treatment effect in any subgroup of patients with $W=w$ (see Appendix B).  Under the null, Assumptions (C1)-(C3) imply that $S\supone \mid W=w$ has the same distribution as $S\supzero \mid W=w$ for all $w$ in the support of $W$, and thus, $\widetilde{\Delta}_H=0.$ Therefore, we may formally define our test statistic for $H_0$ based on the early average treatment effect as
$Z_{H} = \sqrt{n} \widehat{\Delta}_{H}/\widehat{\sigma}_{H},$
where $\widehat{\Delta}_{H}$ is a root-$n$ consistent estimate of $\widetilde{\Delta}_{H}$ and $\widehat{\sigma}_{H}^2$ is the estimated variance  of $\sqrt{n}(\widehat{\Delta}_{H}-\widetilde{\Delta}_{H}).$  We reject $H_0$  when $|Z_{H} |$ is large. In Section \ref{sub-estimation}, we propose robust procedures to construct $\widehat{\Delta}_{H}$ and $\widehat{\sigma}_{H}$. Obviously, this is a valid test for both the null $H_{0H}: \widetilde{\Delta}_{H} = 0$ and the null $H_0: \Delta=0.$

One important merit of constructing the test statistic based on an estimator of $\widetilde{\Delta}_H$ is that this earlier average treatment effect is smaller than if we used the true conditional expectations within each treatment group in probability. That is,  $P(\widetilde{\Delta}_{H}\le \Delta)\approx 1$ and thus, $\widetilde{\Delta}_H$ is a conservative measure of the average treatment effect, $\Delta$. Importantly, this early treatment effect and associated test account for heterogeneity in the utility of the surrogate by explicitly utilizing a condition mean function that depends on $W$. In the following section we describe other tests that may be considered; in our numerical studies, we compare our approach with these alternatives.

\subsection{Alternative Testing Approaches}
We consider two alternative tests that would be reasonable options for testing $H_0$ in this setting. The first quite obvious approach is simply to assume the primary outcome is measured in the current study and use primary outcome information to estimate $\Delta$ and conduct a t-test of $H_0:\Delta=0$. This reflects the gold standard as it directly tests the hypothesis we are interested in. Importantly though, the whole point of this setting is to provide a way to \textit{not} have to measure the primary outcome. We include this option so that we can compare to this gold standard.

The second alternative test we examine is one which uses information from the prior study about the relationship between the surrogate and the primary outcome, but does not account for heterogeneity. This test is an extension of a test proposed in Parast et al. (2019)\cite{parast2019using} which was developed for the time-to-event outcome setting. Our description of it here, for a non-survival setting, is new and will be useful in practice for those analyzing a non-survival study in a setting with no heterogeneity in the utility of the surrogate. Similar to our proposed test, but without regard for $W$, we note that $ \Delta = \int \mu_1(s)dF \supone(s) - \int \mu_0(s)dF \supzero(s)$
where $\mu_g(s) = E( Y \supg | S \supg)$ which motivates the following \textit{earlier} treatment effect definition:
\begin{eqnarray*}
\Delta_P &=& \int \mu_0(s) dF \supone(s) - \int \mu_0(s) dF \supzero(s) = \int \mu^p_0(s) dF \supone(s) - \int \mu^p_0(s) dF \supzero(s)
\end{eqnarray*}
where $\mu^p_0(s) \equiv E(Y \supzerop = y | S \supzerop = s)$.
Since $\mu_0^p(s)$ is unknown, we approximate $\Delta_P$ with
\begin{eqnarray*}
\widetilde{\Delta}_P &=& \int \widehat{\mu}_0(s) dF \supone(s) - \int \widehat{\mu}_0(s) dF \supzero(s) = \int \widehat{\mu}^p_0(s) dF \supone(s) - \int \widehat{\mu}^p_0(s) dF \supzero(s).
\end{eqnarray*}
where $\widehat{\mu}_p^0(s)$ is a consistent estimator of $\mu_0^p(s).$
As with the proposed test, this early treatment effect quantity replaces $\mu_g(s)$ with $\widehat{\mu}_0(s)$ for both treatment groups and will ensure it is a lower bound on the $\Delta$ under certain conditions. This test, however, requires the assumption that $\widehat{\mu}_0^p(s)\approx \mu^p_0(s) = \mu_0(s)$ i.e., that this conditional expectation in the control group is the same in the current study as the prior study. It is important to note that this assumption may not hold when there is heterogeneity in the utility of the surrogate marker. To test $H_0: \Delta = 0$, we instead test $H_{0P}: \widetilde{\Delta}_{P} = 0$ and define the test statistic for $H_{0P}$ based on the early treatment effect as
$Z_{P} = \sqrt{n} \widehat{\Delta}_{P}/\widehat{\sigma}_{P},$
where $\widehat{\Delta}_{P}$ is a root-$n$ consistent estimate of $\widetilde{\Delta}_{P}$ and $\widehat{\sigma}_{P}^2$ is the estimated variance  of $\sqrt{n}(\widehat{\Delta}_{P}-\widetilde{\Delta}_{P}).$  We reject $H_{0P}$ (and $H_0$)  when $|Z_{P} |$ is large.

In Appendix C, we discuss estimation and testing for $\Delta$ using the primary outcome, propose estimation procedures to obtain $\widehat{\Delta}_{P}$ and $\widehat{\sigma}_{P}$, and discuss why we do not consider directly testing the surrogate. Intuitively, we would expect that both our proposed test and this test based on $\widetilde{\Delta}_P$ should work well when there is no heterogeneity. When there is heterogeneity, we expect that the test based on $\widetilde{\Delta}_P$ (or even $\Delta_P$) could lead to erroneous conclusions about the treatment effect and/or have less power than the proposed test.

\section{Estimation and Inference} \label{sub-estimation}
\subsection{Estimation of Proposed $\widetilde{\Delta}_H$ }\label{sub-estimation-delta}
For our proposed testing procedure, we first define
$$\widehat{\mu}_{0}^p(s, w)= \frac{\sum_{i=1}^{n_0^p}K_{h_2}(S^p_{0i}-s)K_{h_3}(W^p_{0i}-w)Y^p_{0i}}{\sum_{i=1}^{n_0^p}K_{h_2}(S^p_{0i}-s)K_{h_3}(W^p_{0i}-w)}, ~\mbox{and}$$
$$\widehat{m}_g(w; \mu(\cdot, \cdot))=\frac{\sum_{i=1}^{n_g}K_{h_g}(W_{gi}-w){ \mu(S_{gi}, W_{gi})}}{\sum_{i=1}^{n_g}K_{h_g}(W_{gi}-w)},$$
as nonparametric smoothed estimators of the conditional expectation of $Y^{(0)}$ given $(S^{(0)}, W)=(s, w)$ in the prior study, and the conditional expectation of $\mu(S^{(g)}, W)$ given $W=w$ and a bivariate function $\mu(\cdot, \cdot)$ in the current study, respectively. Here, $K_h(\cdot) = K(\cdot/h)/h$, $K(\cdot)$ is a smooth symmetric density function with finite support, $h_0, h_1, h_2, h_3$ are specified bandwidths which may be data dependent, and $n_{0}^p$ denotes the sample size of group $Z=0$ in the prior study. We utilize undersmoothing and select all bandwidths throughout to be of order $O(n^{-\epsilon}), \epsilon\in (1/4, 1/2),$ to eliminate the asymptotic bias,  where $n=n_1+n_0$ in an effort to avoid a need for bias correction in subsequent statistical inference.

A very straightforward estimate of $\widetilde{\Delta}_H$ would be
\begin{equation}
n_1^{-1} \sum_{i=1}^{n_1} \widehat{\mu}_{0}^{(p)}(S_{1i}, W_{1i}) - n_0^{-1} \sum_{i=1}^{n_0}\widehat{\mu}_{0}^{(p)}(S_{0i}, W_{0i})\label{simple_est}
\end{equation}
which simply takes our estimated conditional mean function from the prior study and applies it to data in the current study. However, it is possible for us to improve upon this estimator in terms of efficiency. To do this, we note that
\begin{eqnarray*}
\widetilde{\Delta}_H&=&E\left[E\left(\widehat{\mu}_0^p(S^{(1)}, W) \mid W\right)\right]-E\left[E\left(\widehat{\mu}_0^p(S^{(0)}, W)\mid W\right)\right]\\
        &\approx& E\left[\hat{m}_1(W; \widehat{\mu}_0^p)\right]-E\left[\hat{m}_0(W; \widehat{\mu}_0^p)\right],
\end{eqnarray*}
and thus we now consider an estimate of $\widetilde{\Delta}_H$ as
\begin{equation}
n_1^{-1} \sum_{i=1}^{n_1} \widehat{m}_{1}(W_{1i}; \widehat{\mu}_0^p ) - n_0^{-1} \sum_{i=1}^{n_0} \widehat{m}_{0}(W_{0i}; \widehat{\mu}_0^p), \label{simple_est2}
\end{equation}
which is asymptotically equivalent to (\ref{simple_est}). Note that this estimate only uses $S^{(g)}$ and $W$ data from the current study (no $Y$ data from the current study) and $\widehat{\mu}_0^p(s, w)$, which in turns depends on
$S^{(0p)},W^p,Y^{(0p)}$ data in group $Z=0$ from the previous study.

While either (\ref{simple_est}) or (\ref{simple_est2}) would be consistent estimates of $\widetilde{\Delta}_H$, we utilize the fact that the distributions of $W$ from the two treatment arms are identical due to randomization and construct the estimator:
\begin{equation}
 \widehat{\Delta}_{H} = \frac{1}{n_1+n_0}\left \{ \left[\sum_{i=1}^{n_0} \widehat{m}_1(W_{0i}; \widehat{\mu}_0^p)+\sum_{i=1}^{n_1}\widehat{m}_1(W_{1i}; \widehat{\mu}_0^p)\right]-\left[\sum_{i=1}^{n_0} \widehat{m}_0(W_{0i}; \widehat{\mu}_0^p) + \sum_{i=1}^{n_1}\widehat{m}_0(W_{1i}; \widehat{\mu}_0^p)\right] \right \}. \label{ourest}
 \end{equation}
We show in Appendix D that (\ref{ourest}) improves upon the efficiency of (\ref{simple_est2}).  Essentially,  $\widehat{\Delta}_H$ is equivalent to an augmented version of the simple estimator (described below), taking advantage of the independence of $W$ and treatment, since treatment was randomized.

 In Appendix D we show that conditional on $\widehat{\mu}_0^p(\cdot, \cdot)$,   $\widehat{\Delta}_{H}$ is a consistent estimate of $\widetilde{\Delta}_H,$
and that $\sqrt{n}\{ \widehat{\Delta}_{H} - \widetilde{\Delta}_{H} \}$ weakly converges to  a mean zero normal distribution as $n \rightarrow \infty$. A consistent estimate of the conditional variance of $\widehat{\Delta}_{H}$ given the prior study, $\sigma_{H}^2$, can be obtained as $$\hat{\sigma}_H^2=\frac{1}{n_1^2}\sum_{i=1}^{n_1}\left(\widetilde{S}_{1i}- \pi_0 \hat{m}_1(W_{1i}; \widehat{\mu}_0^p)-\pi_1\hat{m}_0(W_{1i}; \widehat{\mu}_0^p)-\pi_1\widehat{\Delta}_H\right)^2$$
$$+\frac{1}{n_0^2}\sum_{i=1}^{n_0}\left(\widetilde{S}_{0i}- \pi_0 \hat{m}_1(W_{0i}; \widehat{\mu}_0^p)-\pi_1 \hat{m}_0(W_{0i}; \widehat{\mu}_0^p)-\pi_0\widehat{\Delta}_H\right)^2$$
where $\pi_g = n_g/n$ and $\widetilde{S}_{gi}=\mu_0^{(p)}(S_{gi}, W_{gi})$. Our testing procedure uses the test statistic $Z_{H} = \widehat{\Delta}_{H}/\widehat{\sigma}_{H}$ and rejects the null hypothesis when $|Z_H| > \Phi^{-1}(1-\alpha/2)$.  As $n_{0p}\rightarrow \infty$, $\widetilde{\Delta}_H-\Delta_H=o_p(1)$ and $\widetilde{\Delta}_H$ can be viewed as a consistent estimator of $\Delta_H$.  More importantly, under Assumptions (C1), (C2), (C3) and (C5),  $P(\widetilde{\Delta}_H\le \Delta)\rightarrow 1$ as $n \rightarrow \infty,$ indicating that the test for $\widetilde{\Delta}_H=0$ is a valid test for $\Delta=0$ with probability approaching 1 as the sample size of the prior study increases to infinity.\\

\noindent Remark. \textit{The efficiency of the simple estimator
\begin{align*}
&n_1^{-1} \sum_{i=1}^{n_1} \widehat{m}_{1}(W_{1i}; \widehat{\mu}_0^p) - n_0^{-1} \sum_{i=1}^{n_0} \widehat{m}_{0}(W_{0i}; \widehat{\mu}_0^p)\approx  n_1^{-1} \sum_{i=1}^{n_1} \widehat{\mu}_{0}^{(p)}(S_{1i}, W_{1i}) - n_0^{-1} \sum_{i=1}^{n_0}\widehat{\mu}_{0}^{(p)}(S_{0i}, W_{0i}),
\end{align*}
 can be improved by considering the fact that
$ E[m(W_{1i}; \widehat{\mu}_0^p)]=E[m(W_{0i}; \widehat{\mu}_0^p)]$
for any transformation $m(\cdot)$ due to randomization. Specifically, one may consider a new class of consistent estimators indexed by $m(\cdot): R \rightarrow R$,
$$\left\{ n_1^{-1} \sum_{i=1}^{n_1} \left[\widehat{\mu}_{0}^{(p)}(S_{1i}, W_{1i})-m(W_{1i}; \widehat{\mu}_0^p)\right] - n_0^{-1} \sum_{i=1}^{n_0}\left[\widehat{\mu}_{0}^{(p)}(S_{0i}, W_{0i})-m(W_{0i}; \widehat{\mu}_0^p)\right]\right\}.$$
The optimal choice of $m(\cdot)$ minimizing the asymptotic variance is
$$m_{opt}(w)=\pi_0 E(\widehat{\mu}_0^{(p)}(S_1, w)|W_1 =w)+\pi_1 E(\widehat{\mu}_0^{(p)}(S_0, w)|W_0 =w).$$
 In practice,
$m_0(w)$ can be consistently estimated by
$\widehat{m}_{opt}(w)=\pi_0 \widehat{m}_1(w; \widehat{\mu}_0^{(p)})+\pi_1\widehat{m}_0(w; \widehat{\mu}_0^{(p)}).$
Denote the resulting estimator of $\widetilde{\Delta}_H$ by
$$\widehat{\Delta}_H^{AUG}=n_1^{-1} \sum_{i=1}^{n_1} \left[\widehat{\mu}_{0}^{(p)}(S_{1i}, W_{1i})-\widehat{m}_{opt}(W_{1i}; \widehat{\mu}_0^p)\right] - n_0^{-1} \sum_{i=1}^{n_0}\left[\widehat{\mu}_{0}^{(p)}(S_{0i}, W_{0i})-\widehat{m}_{opt}(W_{0i}; \widehat{\mu}_0^p)\right].$$}
 \textit{\noindent In Appendix D we show that conditional on $\widehat{\mu}_0^{(p)}(\cdot, \cdot)$,  $\widehat{\Delta}_{H}^{AUG}$ is a consistent estimate of $\widetilde{\Delta}_H$ and that $\sqrt{n}( \widehat{\Delta}_{H}^{AUG} - \widetilde{\Delta}_{H} ) $ weakly converges to  a mean zero normal distribution as $n \rightarrow \infty$. The conditional variance of $\widehat{\Delta}_H^{AUG}\mid \widehat{\mu}_0^{(p)}(\cdot, \cdot),$ $\sigma_{AUG}^2,$ can be consistently estimated by
$$\widehat{\sigma}_{AUG}^2=\frac{1}{n_1^2}\sum_{i=1}^{n_1} \left[\widehat{\mu}_{0}^{(p)}(S_{1i}, W_{1i})- \widehat{m}_1(W_{1i}; \widehat{\mu}_0^p)\right]^2 +\frac{1}{n_0^2}\sum_{i=1}^{n_0} \left[\widehat{\mu}_{0}^{(p)}(S_{0i}, W_{0i})-\widehat{m}_0(W_{0i}; \widehat{\mu}_0^p)\right]^2$$
$$+\frac{\pi_1^2}{n_1^2}\sum_{i=1}^{n_1}\left[\widehat{m}_1(W_{1i}; \widehat{\mu}_0^p)- \widehat{m}_0(W_{1i}; \widehat{\mu}_0^p)-\widehat{\Delta}_H)\right]^2+\frac{\pi_0^2}{n_0^2}\sum_{i=1}^{n_0}\left[\widehat{m}_1(W_{0i}; \widehat{\mu}_0^p)- \widehat{m}_0(W_{0i}; \widehat{\mu}_0^p)-\widehat{\Delta}_H\right]^2.$$
In Appendix D, we show that $\widehat{\Delta}_H^{AUG}$ is asymptotically equivalent to our proposed $\widehat{\Delta}_H$ and $\widehat{\sigma}_H/\widehat{\sigma}_{AUG}=1+o_p(1).$}

\subsection{Inference}
To construct a confidence interval for $\widetilde{\Delta}_H$ we use our estimated variance $\widehat{\sigma}^2_H$ and define a $100(1-\alpha)\%$ confidence interval as $\widehat{\Delta}_H \pm Z_{1-\alpha/2} \widehat{\sigma}_H$. We examine the empirical performance of our proposed estimation procedure, variance estimation, confidence interval construction, and testing procedure in Section \ref{sims}.

It is important to note that we consider the prior study, the study from which we estimate the conditional mean function, $\widehat{\mu}_{0}^p(s, w)$, as fixed. This is a reasonable assumption given that in practice, there is truly some previously conducted prior study which one is using to inform testing in the current study. However, one could argue that this prior study should be considered random and that all inference should be derived as such. In such a case, the estimation of our point estimate $\widehat{\Delta}_H$ would remain the same but the standard estimation and confidence interval construction would be more complex.

\subsection{Multiple Baseline Covariates} \label{multiple}
While in this paper we focus only on heterogeneity with respect to a single baseline covariate, it may be the case that there is heterogeneity with respect to multiple baseline covariates. In such a case, one still can consider a straightforward estimator for the treatment effect using surrogate marker and baseline covariates:
\begin{equation*}
n_1^{-1} \sum_{i=1}^{n_1} \widehat{\mu}_{0m}^{(p)}(S_{1i}, \bW_{1i}) - n_0^{-1} \sum_{i=1}^{n_0}\widehat{\mu}_{0m}^{(p)}(S_{0i}, \bW_{0i})
\end{equation*}
where $\widehat{\mu}_{0m}^{(p)}(s, \bw)$ is an estimator of $\mu_0(s, \bw) \equiv  E\left(Y^{(0)} \mid S^{(0)} = s, \bW = \bw\right)$ and $\bW$ is a baseline covariate vector of interest (including an intercept term, with a slight abuse of notation). The difficulty is that fully nonparametric estimation of $\mu_0(s, \bw)$ will likely be infeasible for practical sample sizes with a vector $\bW$ of moderate dimension, e.g., $\ge 3.$  In such a case, one may be willing to consider a parametric or semi-parametric model. For example, an estimator can be obtained based on a simple regression model $\mu_0(s,\bw) = g_Y\left(\beta_0s + \bbeta_1'\bw\right),$ where $g_Y(\cdot)$ is a known, strictly increasing link function and $\beta_0$ and $\bbeta_1$ are unknown regression coefficients to be estimated based on the prior study. Alternatively, one could consider a more flexible varying coefficient model for $\mu_{0}^p(s, \bw)$ such as ${\mu}_{0}(s, \bw) = g_Y\{\bB(s)'\bw\},$ where $\bB(s) = \{\bbeta_1(s), \bbeta_2(s),..., \bbeta_L(s) \}'$, and $\bbeta_l(s)$ is the unknown smooth function of $s$ to be estimated nonparametrically. This modeling approach would allow complex interactions between $S$ and $\bW$. Here, we use the additional subscript $m$ in $\widehat{\mu}_{0m}^{(p)}(\cdot,\cdot)$ to emphasize the fact that this estimator of $\mu_0(\cdot, \cdot)$ will now be fully or partially dependent on model assumptions, i.e., model-based. Certainly, given this model dependence, robustness (or lack thereof) to model misspecification would need to be carefully considered when using this approach in practice.

\section{Simulation Study} \label{sims}
\subsection{Simulation Goals and Setup}
The two main goals of our simulation study were: 1) to examine the finite sample properties of our estimation procedure for $\widetilde{\Delta}_H$ in terms of bias, accuracy of our variance calculation, and coverage of constructed confidence intervals, and 2) to compare testing results based on the three different testing quantities: $\widehat{\Delta}$ (using the primary outcome, gold standard) vs. $\widehat{\Delta}_P$ (using the surrogate marker, ignoring heterogeneity) vs.  $\widehat{\Delta}_H$ (using the surrogate marker, accounting for heterogeneity). For the testing results, we focus on the point estimates themselves, the resulting effect sizes (point estimate/standard error estimate), and power. {Importantly, when there is heterogeneity, we do not necessarily aim to demonstrate improved power with our proposed approach but rather, to demonstrate settings where the testing procedure using $\widehat{\Delta}_P$ (using the surrogate marker, ignoring heterogeneity) can be incorrect.}

To achieve these goals, we examined eight simulation settings. For all settings, results were summarized over 500 replications; we examined all settings with $(n_1^p, n_0^p)=(1000, 800)$ (sample sizes in prior study) and $(n_1, n_0)=(300, 300)$ (sample sizes in current study). All simulation settings were also repeated with $(n_1^p, n_0^p)=(300, 300)$ (sample sizes in prior study) and $(n_1, n_0)=(300, 300)$; results were similar and are not shown here. In setting 1, we generated data such that there was heterogeneity in the utility of the surrogate with respect to a baseline covariate \textit{and} the distribution of this baseline covariate was different in the current study compared to the prior study. Specifically, in the prior study, which is fixed in all simulations,  $W_{1i}^p \sim U(0,10)$, $W_{0i}^p \sim U(0,10)$, $S_{1i}^p \sim gamma(shape = 2.78, scale =2.78),$ and $S_{0i}^p \sim gamma (shape = 2.5, scale = 2.5).$ We then generate the outcomes from:

\begin{align*}
Y_{1i}^p & = I(W_{1i}^p < 5)(3.5 + 5S_{1i}^p) + I(W_{1i}^p \ge 5)(16   S_{1i}^p)    + N(0,16), \\
Y_{0i}^p & = I(W_{0i}^p < 5)(3.2 + 4S_{0i}^p) + I(W_{0i}^p \ge 5)(15.95S_{0i}^p)    + N(0,16).
\end{align*}

 \noindent where throughout $N(a,b)$ indicates a normal distribution with mean $a$ and variance $b$. The motivation behind this setup was (a) to generate a surrogate marker where higher values are desirable and the surrogate level tends to be higher in the treated group, and (b) to generate an outcome where the surrogate marker is positively associated with the outcome but this association is stronger in magnitude in the treated group, reflecting residual treatment effect beyond the surrogate marker. In addition, to induce heterogeneity, we generate data such that the treatment effect on the primary outcome and the association between primary outcome and surrogate marker depend on whether the covariate is less than or greater than 5.  With this setup, there was a statistically significant heterogeneity in surrogacy based on the test for heterogeneity proposed by Parast et al. (2021); the estimated proportion of treatment effect explained by the surrogate marker was 0.52 for $W_{gi}^p <5$ and 0.95 for $W_{gi}^p \geq 5, g\in \{0, 1\}.$ In this setting, the $(S_{gi}, Y_{gi})\mid W_{gi}$ in the current study was generated the same as in the prior study, but $W_{1i}$ and $W_{0i}$ were generated from a $U(0,4)$, which is different from the prior study. Note that for all patients in the current study, the surrogate strength is not very strong and thus, we would expect that using the surrogate but ignoring heterogeneity will lead to an overestimation of the treatment effect. While the variability of the primary outcome, $Y_{gi}$, is large in both treatment groups, the size of the treatment effect is large as well. For example, in this setting, our results will show that the average estimated treatment effect on the outcome in the current study is 14.10, and the empirical power of testing the treatment effect is 100\% using the primary outcome only.

 In setting 2, $W_{gi}^p$ and $Y_{gi}^p|S_{gi}^p, W_{gi}^p$ in the prior study were generated exactly the same as in setting 1, but $S_{1i}^p \sim gamma(shape = 2.66, scale =2.66)$ and $S_{0i}^p \sim gamma (shape = 2.5, scale = 2.5).$ The motivation behind this change in the distributions for the surrogate marker is that we aimed to make the treatment effect on both the primary outcome and surrogate marker smaller than in setting 1, in order to explore how the various tests performed when less power would be expected. As in setting 1, there was significant heterogeneity in surrogacy with the estimated proportion of treatment effect explained by the surrogate being 0.39 for $W_{gi}^p<5$ and 0.90 for $W_{gi}^p\geq 5$. The current study was generated the same as the prior study except that $W_{1i}$ and $W_{0i}$ were generated from a $U(6,10)$ distribution. In contrast to setting 1, for all patients in the current study, the surrogate is strong and thus, we would expect that using the surrogate but ignoring heterogeneity will lead to an underestimation of the treatment effect. With respect to the size of the treatment effect and empirical power in this setting, our results will show that the average treatment effect on the outcome in the current study is 13.34 , and the empirical power of testing the treatment effect is 69\% using the primary outcome only. 

 In setting 3, $(W_{gi}, S_{gi})$ in the prior study were generated as in setting 2, but  $Y_{1i}^p  = I(W_{1i}^p < 5)(3.5 + 5\times7) + I(W_{1i}^p \geq 5)(16S_{1i}^p)+ N(0,16)$ and $Y_{0i}^p = I(W_{0i}^p < 5)(3.2+4\times6.25) + I(W_{0i}^p \ge 5)(15.95S_{0i}) + N(0,16)$. The motivation behind this change in the distributions for $Y$ was to explicitly make the surrogate useless among those with $W_{gi}^p<5$ i.e., a more extreme version of setting 2. As expected, there was significant surrogacy heterogeneity with the treatment effect on the surrogate marker not explaining any of the treatment effect on the primary outcome among patients with $W_{gi}^p<5$, and explaining the majority of the treatment effect on the primary outcome among patients with $W_{gi}^p\geq 5$ (proportion explained $\approx$ 0.92).  Similar to setting 2, the current study was generated the same as the prior study except that $W_{1i}$ and $W_{0i}$ were generated from a $U(6,10)$ distribution and thus, we expect a potentially larger gain in power using our proposed approach (though again, this is not our primary goal). With respect to the size of the treatment effect and empirical power in this setting, our results will show that the average treatment effect on the primary outcome in the current study is 13.34 , and the empirical power of testing the treatment effect is 69\% using the primary outcome only, parallel to setting 2. 

In setting 4, the prior study was generated exactly the same as in setting 1, and the current study was generated exactly the same as the prior study, i.e., $W_{1i}$ and $W_{0i}$ were generated from a $U(0,10)$ distribution. Here, even though there is heterogeneity as described above for setting 1, since the covariate distribution is the same in prior and current studies, we expect the tests ignoring vs. accounting for heterogeneity to produce similar results. With respect to the size of the treatment effect and empirical power in this setting, our results will show that the average treatment effect on the primary outcome in the current study is 19.12 , and the empirical power of testing the treatment effect is 96\% using the primary outcome only. 

In setting 5, data were generated such that there is no heterogeneity. Specifically, in the prior study, $W_{1i}^p \sim U(0,10)$, $W_{0i}^p \sim U(0,10)$, $S_{1i}^p \sim gamma(shape = 2.78, scale =2.78)$, $S_{0i}^p \sim gamma (shape = 2.5, scale = 2.5)$, $Y_{1i}^p = 3.5 + 5S_{1i}^p + N(0,1),$ and $Y_{0i}^p = 3.2+4S_{0i}^p + N(0,1),$ independent of the baseline covariate.  The proportion of the treatment effect explained by the surrogate in the prior study was 0.47, which is homogeneous in the study population. Data from the current study was distributed the same as for the prior study. The purpose of this setting was to examine how the tests perform when there is no heterogeneity and no difference in distribution from the prior study to the current study. With respect to the size of the treatment effect and empirical power in this setting, our results will show that the average treatment effect on the outcome in the current study is 13.90 , and the empirical power of testing the treatment effect is 100\% using the primary outcome only.

In setting 6, data are generated similar to setting 1 but with lower variability in the primary outcome resulting in a much larger effect size. In the prior study, $W_{1i}^p \sim U(0,10)$, $W_{0i}^p \sim U(0,10)$, $S_{1i}^p \sim gamma(shape = 3, scale =3),$ $S_{0i}^p \sim gamma (shape = 2.1, scale = 2.2).$ For $W_{1i}^p < 5$ and $W_{0i}^p<5,$ $Y_{1i}^p = 3.5 + 5S_{1i}^p + N(0,1),$ and $Y_{0i}^p = 1+3S_{0i}^p + N(0,1),$ respectively.  For $W_{1i}^p \geq 5$ and $W_{0i}^p\geq 5,$ $Y_{1i}^p = 16S_{1i}^p + N(0,1)$ and $Y_{0i}^p = 15.8S_{0i}^p + N(0,1),$ respectively.  There was a substantial heterogeneity in the utility of the surrogate with the proportion of treatment effect explained by the surrogate being 0.67 for $W_{gi}^p<5$ and 0.98 for $W_{gi}^p \geq 5$. In the current study, the $S$ and $Y$ were generated the same as in the prior study, but $W_{1i}$ and $W_{0i}$ were generated from a $U(0,4)$ distribution. As in setting 1, since the surrogate strength is not very strong in the current study, we would expect that using the surrogate but ignoring heterogeneity will lead to an overestimation of the treatment effect. With respect to the size of the treatment effect and empirical power in this setting, our results will show that the average treatment effect on the outcome in the current study is 33.70 , and the empirical power of testing the treatment effect is 100\% using the primary outcome only.

Settings 7 and 8 reflect a null treatment effect setting and we include them so that we may examine the empirical Type 1 error rate. In both settings, data from the prior study are generated as $W_{gi}^p \sim U(0,10)$, $S_{gi}^p \sim gamma(shape = 2.5, scale =2.5)$, and $Y_{gi}^p = 3.2 + 4S_{gi}^p + N(0,16)$ for $g=0,1$. That is, there is neither treatment effect on the surrogate marker nor the treatment effect on the primary outcome, and $S_{gi}$ and $Y_{gi}$ are positively associated. In setting 7, data in the current study are generated exactly as the prior study. In setting 8, data in the current study are generated such that $(S_{gi}, Y_{gi})|W_{gi}$ are generated the same as the prior study, but $W_{gi} \sim U(0,4), g\in \{0, 1\},$ i.e., the distribution of the baseline covariate is different in the current study. The purpose of setting 8 is to specifically examine estimation and testing when there is no treatment effect and no heterogeneity, but the current study does have a different patient population compared to the prior study. In both settings, the true treatment effect on the primary outcome is 0 and the empirical Type 1 error of the test using the primary outcome is 0.06. In both settings, there is no empirical evidence that $S$ is an ``informative'' surrogate marker, and no empirical evidence of heterogeneity in surrogacy, as expected.

With respect to our bandwidth selection,  we let $h_0=1.06\times \min(\sigma_{W_0}, IQR_0/1.34) n_0^{-2/5} $ and $h_1=1.06\times \min(\sigma_{W_1}, IQR_0/1.34) n_1^{-2/5} $ where $\sigma_{W_g}$ and $IQR_g$ were the empirical standard deviation and inter-quartile range of $W_g,$ and $h_2= 2\times1.06\times \min(\sigma_{S_0^p}, IQR_1/1.34) n_{0^p}^{-2/5}$ and $h_3= 2\times1.06\times \min(\sigma_{W_0^p}, IQR_2/1.34) n_{0^p}^{-2/5}$ where $\sigma_{S_{0^p}}$ and $IQR_1$ were the empirical standard deviation and inter-quartile range of $S_{0^p},$ respectively, and $\sigma_{W_{0^p}}$ and $IQR_2$ were the empirical standard deviation and inter-quartile range of $W_{0^p}$, and $h_4 = 1.06\times \min(\sigma_{S_0^p}, IQR_1/1.34) n_{0^p}^{-0.31}$. \cite{scott1992multivariate,parast2019using}

\subsection{Simulation Results}
Table \ref{tab1} shows estimation results for $\widehat{\Delta}_H$ for all settings, using our proposed estimating procedure. We examine bias in coverage with respect to both $\widetilde{\Delta}_H$ (fixed prior study) and $\Delta_H$. These results demonstrate good performance with minimal bias, average standard error estimates that are close to the empirical standard error, and coverage of the confidence intervals close to the nominal value of 95\%.

Table \ref{tab2} shows results from testing using $\widehat{\Delta}$, $\widehat{\Delta}_P$, and $\widehat{\Delta}_H$. In setting 1 where there is heterogeneity and the distribution of $W$ in the current study is different from the prior study, results show that $\widehat{\Delta}_P$ overestimates the treatment effect and thus, does not retain the lower boundedness property. In contrast, our approach using $\widehat{\Delta}_H$ does not overestimate the treatment effect. The power using $\widehat{\Delta}_H$ is smaller than that using $\widehat{\Delta}$, but this is expected since the data generation in this setting is such that the population in the current study is composed largely of individuals where the surrogate marker is not very strong. In setting 2 where there is again heterogeneity and the distribution of $W$ in the current study is different from the prior study, results show that both $\widehat{\Delta}_P$ and $\widehat{\Delta}_H$ are less than $\widehat{\Delta}$, but $\widehat{\Delta}_H$ is much closer to $\widehat{\Delta}$ and has power equivalent to that using $\widehat{\Delta}$. This, again, is what was expected since the data generation in this setting is such that the population in the current study is composed largely of individuals where the surrogate marker is strong. {{In setting 3, which is similar to setting 2 but we have made the data more extreme with the surrogate being useless for those with $W<5$, results show a larger departure in $\widehat{\Delta}_P$ from $\widehat{\Delta}$, and a larger decrease in power for  $\widehat{\Delta}_P$ compared to $\widehat{\Delta}_H$.}} In setting 4 where there is heterogeneity but the distribution of $W$ in both the prior study and the current study is the same, we see similar point estimates for $\widehat{\Delta}_P$ and $\widehat{\Delta}_H$ but a slightly higher standard error and lower power for $\widehat{\Delta}_H$. This indicates that in some settings, we may pay a price in terms of power and efficiency when we use the approach that accounts for heterogeneity when it is not necessary. In setting 5, where there is no heterogeneity, we see similar performance for $\widehat{\Delta}_P$ and $\widehat{\Delta}_H$. In setting 6, where we have a very large treatment effect on the primary outcome, there is heterogeneity and the distribution of $W$ in the current study is different from the prior study, results show that, as expected, $\widehat{\Delta}_P$ overestimates the treatment effect and does not retain the lower boundedness property, as in setting 1. In settings 7 and 8, where there is no treatment effect, results show that all three testing procedures perform well with an estimated treatment effect close to zero and Type 1 error rate close to 0.05. We additionally examined the efficiency gain comparing our proposed estimator to the simple estimator in (\ref{simple_est});  indeed, we did observe efficiency gains using our proposed estimator, quantified by the ratio of the estimated standard error using our proposed estimate to that using the simple estimate, that ranged from 0.79-0.98 across settings.

In summary, results from this simulation study show 1) good finite sample performance of our estimation and inference procedures for $\Delta_H$, 2) a potential slight loss in power when using the proposed $\widehat{\Delta}_H$ compared to $\widehat{\Delta}_P$ when accounting for heterogeneity is not needed, and 3) a potential for inaccurate conclusions and/or loss in power when $\widehat{\Delta}_P$ is used instead of the proposed  $\widehat{\Delta}_H$ when accounting for heterogeneity is needed.

\section{Application}
\begin{figure}[htbp]
\begin{center}
\includegraphics[scale=0.95]{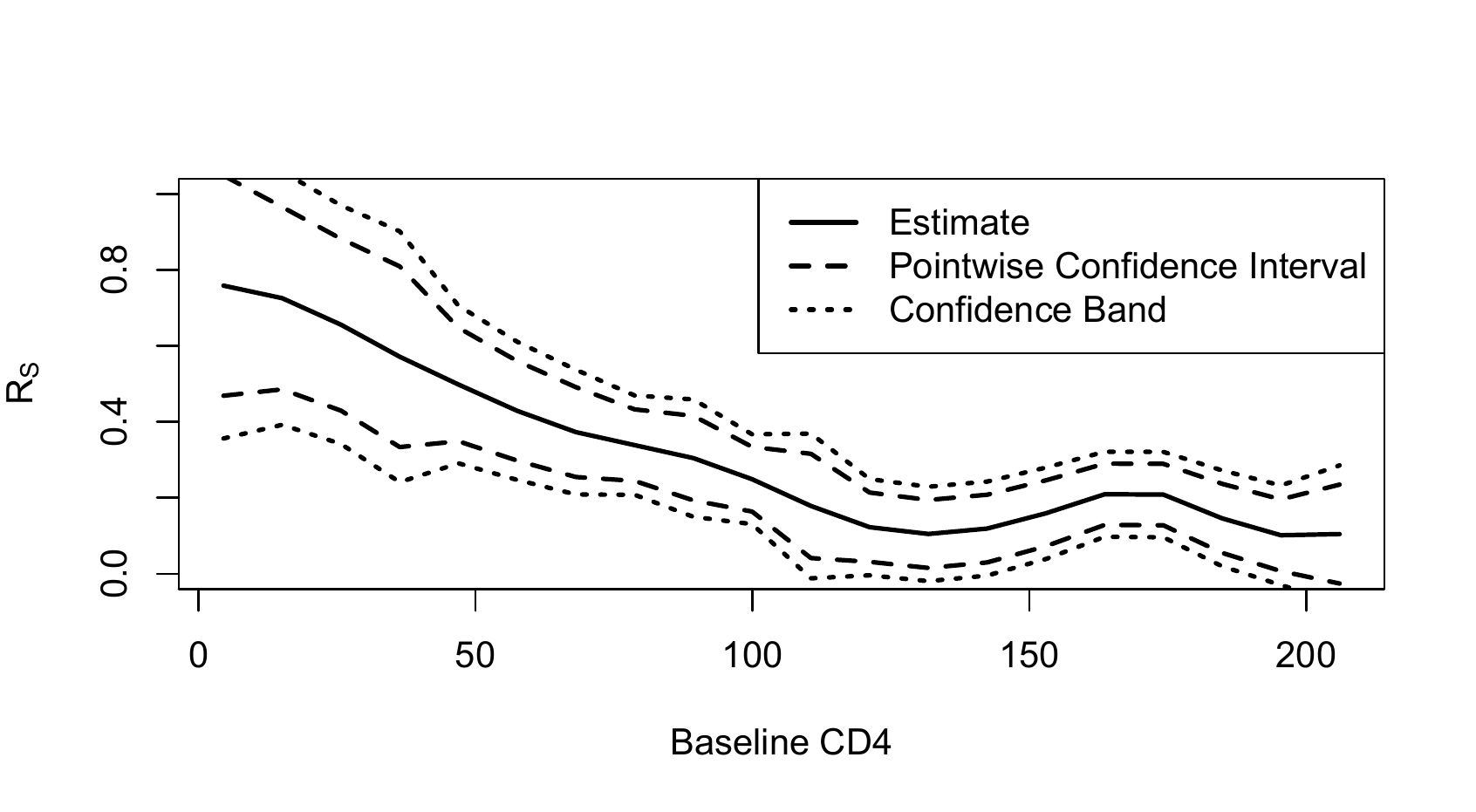}
\end{center}
\caption{Estimated proportion of the treatment effect on the primary outcome (change in RNA) explained by the treatment effect on the surrogate marker (change in CD4), denoted as $R_S$, as a function of baseline CD4}\label{AIDS_figure1}

\begin{center}
\includegraphics[scale=0.95]{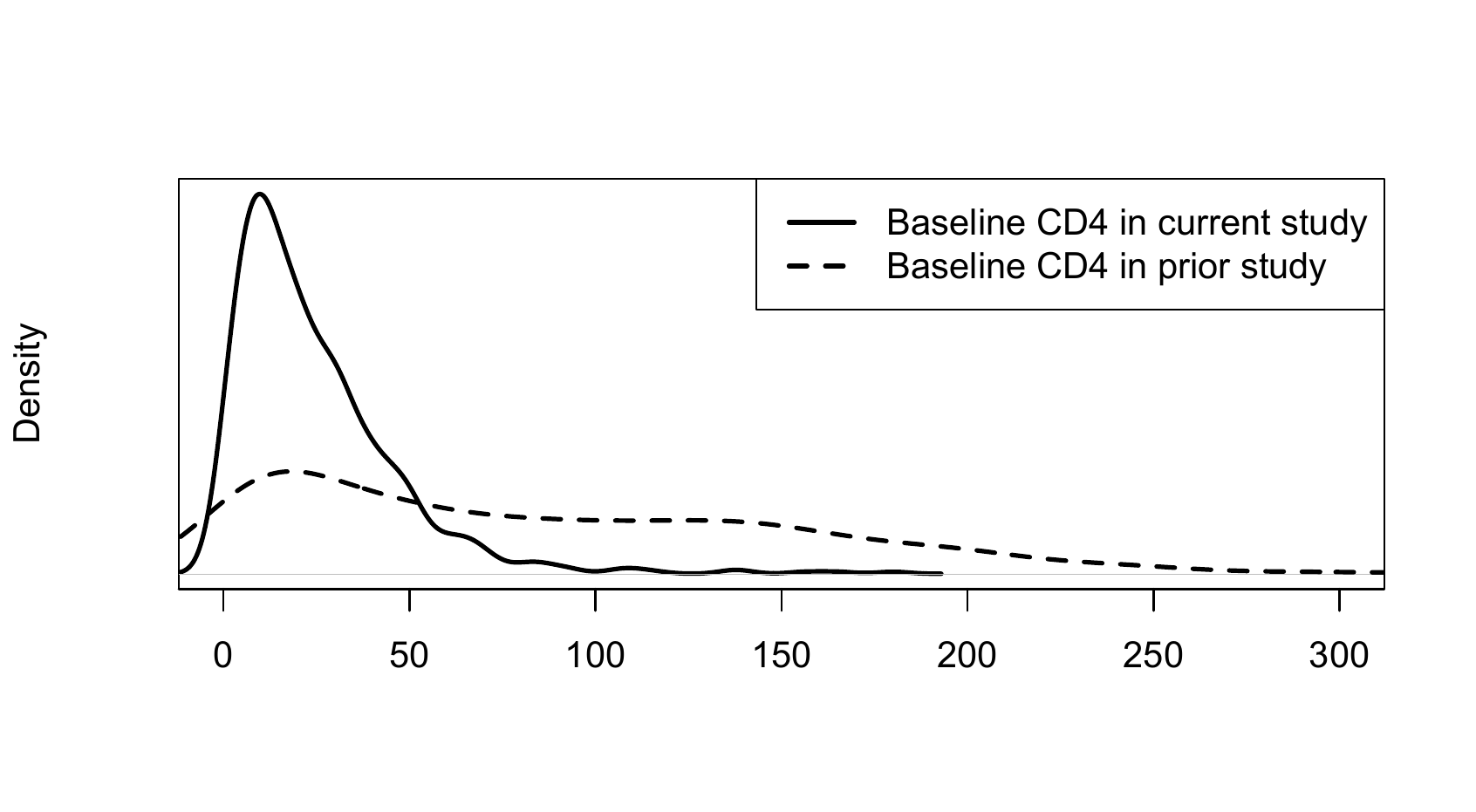}
\end{center}
\caption{Distribution of baseline CD4 in current study vs. prior study}\label{AIDS_figure2}
\end{figure}

We apply our proposed approach to test for a treatment effect based on a heterogeneous surrogate using data from two distinct AIDS clinical trials, the AIDS Clinical Trials Group (ACTG) 320 Study and the ACTG 193A Study. \cite{henry1998randomized,hammer1997controlled} These data are publicly available upon request from the AIDS Clinical Trial Group \cite{ACTG}. We consider the ACTG 320 Study as our prior study and the ACTG 193A Study as our current study. The ACTG 320 study was conducted among HIV-infected patients with a CD4 cell count of 200 or less per cubic millimeter and was a randomized, double-blind trial that compared a two-drug regimen (two nucleoside reverse transcriptase inhibitors [NRTI]) with a three-drug regimen (two NRTIs plus indinavir). There were a total of 830 participants, with 412 in the two-drug regimen group and 418 in the three-drug regimen group. The ACTG 193A study was a randomized, double-blind trial conducted among HIV-infected patients with a CD4 cell count of 50 or less per cubic millimeter. We focus on the comparison of a two-drug regimen (NRTIs) with a three-drug regimen (two NRTIs plus nevirapine). There were a total of 657 participants, with 327 in the two-drug regimen group and 330 in the three-drug regimen group. Our primary outcome $Y$ is the change in plasma HIV-1 RNA from baseline to 24 weeks; our surrogate marker $S$ is change in CD4 cell count from baseline to 24 weeks, as CD4 is relatively less expensive to measure compared to RNA.\cite{calmy2007hiv} Both $Y$ and $S$ are available in ACTG 320 while only $S$ is available in the publicly available data of ACTG 193A. Previous work has demonstrated significant heterogeneity in the utility of $S$ with respect to $W$, baseline CD4 count, with the surrogate strength being stronger among those with a lower baseline CD4 count and weaker among those with a higher baseline CD4 count\cite{parast2021het} as shown in Figure \ref{AIDS_figure1}. We aim to use our proposed method to test for a treatment effect on RNA using CD4 count as a surrogate marker, accounting for the known heterogeneity in the utility of the surrogate which was demonstrated in the prior study.

 In Figure \ref{AIDS_figure2} we show the distribution of the baseline covariate, baseline CD4, in the prior study compared to the current study. Clearly, the current study is composed of a different participant population with lower CD4 counts due to the study eligibility criteria. In Figure \ref{AIDS_figure1}, we also see that the surrogate is strongest in this subgroup. Using our proposed approach, we obtain a treatment effect estimate of $\widehat{\Delta}_H = -0.10$ (standard error [SE] $= 0.03$) with a p-value $< 0.001$. Note that since \textit{lower} plasma HIV-1 RNA is better, a negative change in RNA indicates a beneficial treatment effect for the three-drug regimen. Using the approach that does not account for heterogeneity, we obtain a treatment effect estimate closer to the null, but still significant: $\widehat{\Delta}_P = -0.07 (SE = 0.02), p<0.001$. That is, while the overall conclusion regarding the treatment effect based on the surrogate would be significant using either test, our proposed test provides a treatment effect point estimate that is larger in magnitude. This is expected since the surrogate strength is greater in this subgroup that makes up the current study, and our proposed approach takes advantage of that information.

\section{Discussion}
For settings where it is known that the strength of a surrogate marker varies by a certain baseline characteristic, we have proposed an approach and estimation procedures to appropriately test for a treatment effect using only the surrogate marker, accounting for this known heterogeneity. We demonstrated good finite sample performance of our estimation procedure and showed that our proposed testing procedure can outperform an approach that does not account for heterogeneity.   An R package implementing the methods proposed here, named \texttt{hettest}, is available at \url{https://github.com/laylaparast/hettest}.

While we largely focus, specifically in the numerical studies, on settings where the distribution of $W$ is different in the current study as compared to the prior study, it is still possible for a test based on $\widehat{\Delta}_P$, i.e., ignoring heterogeneity, to provide inaccurate results about the treatment effect when there is heterogeneity in the utility of the surrogate and the $W$ is distributed the \textit{same} in the two studies; we provide an example in Appendix E.

In the presence of heterogeneity, both the treatment effect and the utility of the surrogate marker may depend on $W$.
While we focus exclusively on the average treatment effect in this paper, it may be of interest to test for a treatment effect
based on alternative summaries that account for such heterogeneity. For example, one may define $\Delta_w = E(Y^{(1)} \mid W^{(1)} =w) - E(Y^{(0)} \mid W^{(0)}  = w)$
and the subgroup specific earlier treatment effect $\Delta_H(w) = \int \mu_0^p(s,w)dF^{(1)} (s|w) - \int \mu_0^p(s,w) dF^{(0)}(s|w)$. Then we may test for a treatment effect based on $S$ by examining a functional of $\Delta_H(w)$ such as $\sup_w \Delta_H(w)$ or $\int \Delta_H(w)dw,$ the area under the curve produced by $\Delta_H(w)$. Such alternative summaries of the treatment effect across a baseline covariate, $W$, are not unique to the surrogate marker setting as they have been extensively discussed in the general heterogeneous treatment effect literature. \cite{cai2011analysis,zhao2013effectively} However, these alternative summaries may also prove useful in the heterogeneous surrogate setting and may offer new insights over simply looking at the average treatment effect.

Importantly, we require Assumptions $(C1)-(C4)$ and in practice, they may be violated.  Specifically, if the existing heterogeneity is such that the surrogate is not strong or, worse, the treatment effect on the surrogate marker and primary endpoint may be in different directions for some $w,$ the surrogate should \textit{not} be used as a replacement of the primary outcome for \textit{all} individuals in a future study. Instead, one may consider using the surrogate as a replacement only among those with a $w$ where assumptions $(C1)-(C4)$ hold. To achieve this, one could consider first identifying a region of interest where the surrogacy is sufficiently strong e.g., $\Omega_w$ such that the conditional average treatment effect on the primary endpoint $\Delta(w)\ge \delta_0>0$ and the proportion explained by the surrogate for $W=w$, $R_S(w)=\Delta_H(w)/\Delta(w)$, is between 0.50 and 1.0, and then apply the proposed testing procedure that replaces $Y$ with $S$ for testing the average treatment effect in the subpopulation $\Omega_w$. If one is interested in studying the average treatment effect in the entire study population, one may combine the proposed test statistic with a new but simple test statistic measuring the strength of the treatment effect based on actual primary endpoints $Y$ for patients in the complement of $\Omega$. Such a hybrid approach has the potential to reduce costs if $S$ is less costly to measure than $Y$ and/or reduce the follow-up time needed for those in $\Omega_w$ if $S$ is measured earlier than $Y$. Though not exactly within this context, previous work has explored the potential for auxiliary information (including but not limited to surrogate markers) to improve efficiency when testing for a treatment or intervention effect.\cite{fleming1994surrogate,pepe1992inference} While this is beyond the scope of this paper, further work on this topic within the framework of a heterogeneous surrogate is warranted.


Our proposed approach has some limitations. First, if the current study includes participants with $w$ values outside the observed distribution in the prior study, our approach will not be able to obtain $\widehat{\mu}_{0}^p(s, w)$ for that $w$ without extrapolation. In such a case, when there is observed heterogeneity in the prior study, use of the surrogate marker to test for a treatment effect in the current study should likely be limited to those with $w$ contained in the prior study. Second, given our use of kernel smoothing, we require a relatively large sample size. Robust nonparametric methods for surrogate markers are lacking in general for small sample size settings; future work in this area would be needed. Lastly, we require several assumptions, outlined in Section \ref{assumptions}, which are generally untestable though they may be empirically explored using the observed data. These assumptions are needed for identifiability, to ensure our lower-boundedness property of $\Delta_H$ (i.e., $\Delta_H \leq \Delta$), and to guard against the surrogate paradox which occurs when the surrogate and outcome are positively associated, the treatment has a positive effect on the surrogate, but the treatment in fact has a negative effect on the outcome.\cite{vanderweele2013surrogate} The surrogate paradox is especially of concern here as our primary goal is to make a conclusion about the treatment effect on the primary outcome based on information about the surrogate marker. While these assumptions are strong, they are more likely to hold than the parallel assumptions required for $\Delta_P$\cite{parast2019using} to be valid due to the additional conditioning on $W$. Further work on methods that allow for more relaxed assumptions and/or that allow one to assess sensitivity to violations of these assumptions would be useful.\cite{elliott2015surrogacy}

\section*{Acknowledgements}
Support for this research was provided by National Institutes of Health grant R01DK11835. We are grateful to the AIDS Clinical Trial Group for providing the AIDS clinical trial data.

\bibliographystyle{abbrvnat}
\bibliography{Surrogate_bib}

\clearpage
\begin{table}[hptb]
\begin{center}
\begin{tabular}{|l|c|c|c|c|c|c|c|} \hline
\\ [-1.15em]
 & Estimate& Bias & $\widetilde{\mbox{Bias}}$& ESE& ASE& Cov & $\widetilde{\mbox{Cov}} $ \\ \hline
Setting 1  &  6.32 & 0.07 & 0.05 & 1.82 & 1.79 & 0.96 & 0.96 \\
Setting 2   & 12.53 & 0.05 & 0.07 & 5.39 & 5.22 & 0.94 & 0.94 \\
Setting 3 & 12.52 & 0.05 & 0.07 & 5.39 & 5.22 & 0.94 & 0.94 \\
Setting 4  &14.72 & 0 & 0.05 & 4.12 & 4.13 & 0.96 & 0.95\\
Setting 5  & 5.75 & 0.03 & 0.04 & 1.38 & 1.4 & 0.95 & 0.95 \\
Setting 6  & 12.97 & 0.01 & 0.02 & 1.05 & 1.27 & 0.98 & 0.98 \\
Setting 7  & -0.03 & 0.03 & 0.16 & 1.31 & 1.25 & 0.94 & 0.94 \\
Setting 8  & -0.03 & 0.03 & 0.16 & 1.31 & 1.26 & 0.94 & 0.94 \\
\hline
\end{tabular}
\caption{Estimation results from the simulation study using the proposed procedure to estimate $\widetilde{\Delta}_H$; note that settings 7 and 8 are null settings with no treatment effect; bias and coverage are examined with respect to $\widetilde{\Delta}_H$ (prior study fixed) and $\Delta_H$; $\widetilde{\mbox{Bias}}$ = bias with respect to $\widetilde{\Delta}_H$, quantified as $|\widehat{\Delta}_H - \widetilde{\Delta}_H|/\widetilde{\Delta}_H$ except for settings 7 and 8 where it is quantified without dividing by $\widetilde{\Delta}_H$; Bias = bias with respect to $\Delta_H$,  quantified as $|\widehat{\Delta}_H - \Delta_H|/\Delta_H$ except for settings 7 and 8 where it is quantified without dividing by the truth; ESE = empirical standard error, ASE = average standard error (average of the square root of the closed form variance estimate), $\widetilde{\mbox{Cov}}$ = coverage of 95\% confidence intervals with respect to  $\widetilde{\Delta}_H$; Cov = coverage of 95\% confidence intervals with respect to  $\Delta_H$ \label{tab1}}
\vspace{3mm}
\end{center}
\end{table}

\begin{table}[hptb]
\begin{center}
\begin{tabular}{|l|c|c|c|c|c|} \hline
\multicolumn{6}{|c|}{Setting 1}\\ \hline
 &  Estimate& ESE& ASE& Effect size & Power  \\ \hline
$\Delta$&14.10& 1.64& 1.65& 8.55& 1.00\\
$\Delta_P$&14.53& 3.61& 3.65& 3.99& 0.98\\
$\Delta_H$& 6.32& 1.82& 1.79& 3.62& 0.95\\
\hline
\multicolumn{6}{|c|}{Setting 2}\\ \hline
 &  Estimate& ESE& ASE& Effect size & Power  \\ \hline
$\Delta$&13.34& 5.54& 5.42& 2.47& 0.69\\
$\Delta_P$& 7.64& 3.38& 3.31& 2.31& 0.64\\
$\Delta_H$&12.53& 5.39& 5.22& 2.39& 0.67\\
\hline
\multicolumn{6}{|c|}{Setting 3}\\ \hline
 &  Estimate& ESE& ASE& Effect size & Power  \\ \hline
$\Delta$&13.34& 5.54& 5.42& 2.47& 0.69\\
$\Delta_P$& 6.00& 2.81& 2.76& 2.18& 0.58\\
$\Delta_H$&12.52& 5.39& 5.22& 2.39& 0.67\\
\hline
\multicolumn{6}{|c|}{Setting 4}\\ \hline
 &  Estimate& ESE& ASE& Effect size & Power  \\ \hline
$\Delta$&19.12& 5.17& 5.20& 3.68& 0.96\\
$\Delta_P$&14.64& 3.66& 3.66& 4.01& 0.98\\
$\Delta_H$&14.72& 4.12& 4.13& 3.56& 0.95\\
\hline
\multicolumn{6}{|c|}{Setting 5}\\ \hline
 &  Estimate& ESE& ASE& Effect size & Power  \\ \hline
$\Delta$&13.90& 1.64& 1.65& 8.43& 1.00\\
$\Delta_P$& 5.77& 1.38& 1.38& 4.18& 0.99\\
$\Delta_H$& 5.75& 1.38& 1.40& 4.09& 0.99\\
\hline
\multicolumn{6}{|c|}{Setting 6}\\ \hline
 &  Estimate& ESE& ASE& Effect size & Power  \\ \hline
$\Delta$&33.70& 1.61& 1.60&21.08& 1.00\\
$\Delta_P$&39.12& 3.51& 3.50&11.18& 1.00\\
$\Delta_H$&12.97& 1.05& 1.27&10.23& 1.00\\
\hline
\multicolumn{6}{|c|}{Setting 7}\\ \hline
 &  Estimate& ESE& ASE& Effect size & Type 1 error  \\ \hline
$\Delta$&-0.05& 1.39& 1.35&-0.04& 0.06\\
$\Delta_P$&-0.03& 1.31& 1.27&-0.02& 0.06\\
$\Delta_H$&-0.03& 1.31& 1.25&-0.02& 0.06\\
\hline
\multicolumn{6}{|c|}{Setting 8}\\ \hline
 &  Estimate& ESE& ASE& Effect size & Type 1 error  \\ \hline
$\Delta$&-0.05& 1.37& 1.33&-0.04& 0.06\\
$\Delta_P$&-0.03& 1.31& 1.27&-0.02& 0.06\\
$\Delta_H$&-0.03& 1.31& 1.26&-0.02& 0.06\\
\hline
\end{tabular}
\caption{Testing results from the simulation study comparing testing results based on the three different testing quantities: $\widehat{\Delta}$ (using the primary outcome, gold standard) vs. $\widehat{\Delta}_P$ (using the surrogate marker, ignoring heterogeneity) vs.  $\widehat{\Delta}_H$ (using the surrogate marker, accounting for heterogeneity); ESE = empirical standard error, ASE = average standard error (average of the square root of the closed form variance estimate), Effect size = estimate divided by the estimated standard error (i.e., square root of the closed form variance estimate), Power/Type 1 error = proportion of replications for which the test rejects the null i.e., p-value of the test is $<0.05$ \label{tab2}}
\vspace{3mm}
\end{center}
\end{table}

\clearpage

\clearpage
\appendix

\section*{Appendix A}
\subsection*{Discrete Example}
\begin{figure}[htbp]
\begin{center}
\includegraphics[scale=0.58]{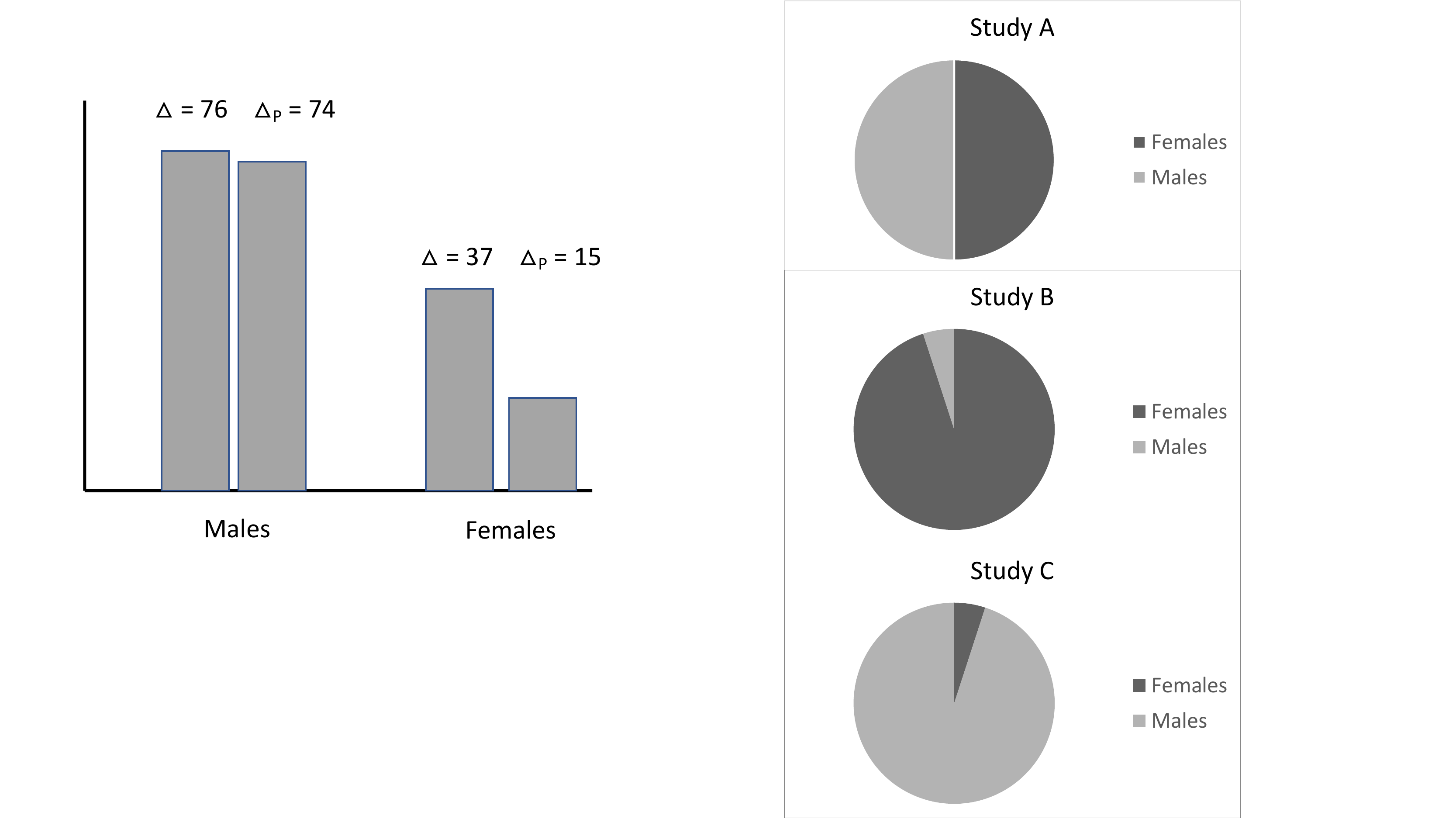}
\end{center}
\caption{Discrete data example}\label{app_fig}
\end{figure}

 Let $Y$ denote the primary outcome and $S$ denote the surrogate marker. 
 We use potential outcomes notation where each person has a potential $\{Y \supone, Y \supzero, S \supone, S \supzero\}$ where $Y \supg$ and $S \supg$ are the outcome and surrogate when the patient receives treatment $g$. 
 Our main quantity of interest is the treatment effect on the primary outcome quantified as $ \Delta \equiv E(Y \supone - Y \supzero) = E(Y \supone) - E(Y \supzero).$ The earlier treatment effect incorporating $S$ information is defined in the main text as
\begin{eqnarray}
\Delta_P &=& \int \mu^p_0(s) dF \supone(s) - \int \mu^p_0(s) dF \supzero(s)
\end{eqnarray}
where $\mu^p_0(s) \equiv E(Y \supzerop = y | S \supzerop = s)$. In this example, we will have heterogeneity in the utility of the surrogate with respect to gender. Consider our prior study, which we refer to as Study A in this example, and is shown in Figure  \ref{app_fig}. The Study A sample is 50\% female and 50\% male. For all individuals, $(S\supone, S\supzero)$ are independent of gender, and $\left\{E(S\supone), E(S\supzero)\right\}=(10, 5).$ For females, $E(Y\supone\mid S\supone=s) = 3+5s$ and $E(Y\supzero \mid S\supzero=s)=1+3S$. It can be shown that for females, $\Delta = 53-16 = 37$ and $\Delta_P = 15.$ The proportion of the treatment effect on the primary outcome that is explained by the surrogate among females is thus 15/37=41\%, which would not be considered as a strong surrogacy. For males, $E(Y\supone\mid S\supone=s) = 15s$ and $E(Y\supzero \mid S\supzero=s) = 14.8S$. It can be shown that for males, $(\Delta, \Delta_P) = (76, 74)$ and the proportion explained by the surrogate marker is 97\% among males, representing strong surrogacy.


To calculate $\Delta_P$ for a future study, let's consider the conditional mean that is central to this calculation, $\mu_0^p(s) = E(Y^{(0p)} = y|S^{(0p)} = s)$ where the superscript $p$ indicates that this is referring to the prior study, i.e., study A.  In this example, this would be $\mu_0^p(s) = 0.5\times(1+3s) + 0.5\times 14.8s = 8.9s+0.5.$  Now assume our current study is Study B shown in Figure \ref{app_fig} which is 95\% female and 5\% male. Importantly, the joint distributions of $(Y\supone, Y\supzero, S\supone, S\supzero)$ in males and females remain as described above; the only difference is the distribution of gender. The treatment effect, $\Delta$ in this new study is $0.95\times 37 + 0.05\times76 = 38.95$. If one were to calculate $\Delta_P$ not accounting for this known heterogeneity in the utility of the surrogate, the quantity obtained would be $\Delta_P = 8.9\times 10 + 0.5 - 8.9\times 5 - 0.5 = 44.5$, recalling that $E(S\supone)=10$ and $E(S\supzero)=5$ for all individuals in both studies. However, using our proposed approach which does account for heterogeneity, we use $\Delta_H$ as the earlier treatment effect, defined in the main text as:
\begin{eqnarray*}
\Delta_H &=&   \int \mu_0^p(s,w)dF^{(1)} (s, w) - \int \mu_0^p(s,w) dF^{(0)}(s, w).
\end{eqnarray*}
Thus, $\Delta_H = 95\%\times (1+3\times 10) + 5\%\times (14.8\times 10) - 95\%\times (1+3\times 5) - 5\%\times (14.8\times 5) = 17.95$. Therefore $\Delta_H < \Delta<\Delta_P$ and $\Delta_P$ no longer retains the property of providing a lower bound on the treatment effect on $Y$.

Now we consider a study, labeled Study C in Figure \ref{app_fig}, which is 95\% males and 5\% females. Using similar calculations, we can show that $\Delta = 74.05$, $\Delta_P = 44.05$ and $\Delta_H = 71.05$. Thus, in this case, $\Delta_H$ will provide better lower bound for $\Delta$ and the test based on $\Delta_H$ is expected to be more powerful than that based on $\Delta_P$. The discrete case, as illustrated in this example, is relatively straightforward in terms of how to go about calculating the needed quantities separately by group and appropriately accounting for the different distribution in the new study. The continuous baseline covariate case, however, is more complex, and our Appendix C presents an example such that even if the prior and current studies have the same distribution for covariates, $\Delta_P$ may still fail to be a valid lower bound for $\Delta.$

\section*{Appendix B}
As noted in this text, Assumptions $(C1)-(C3)$ together guarantee that $E(Y\supone \mid W=w)\ge E( Y \supzero \mid W=w)$, for all $w$ in the support of $W$. This result is due to the derivation:
\begin{align*}
\Delta(w)=&E(Y\supone \mid W=w)- E( Y \supzero \mid W=w)\\
=& \int_s E(Y\supone \mid S\supone=s, W=w)dF\supone(s\mid w)-\int_s  E(Y\supzero \mid S\supzero=s, W=w)dF\supzero(s\mid w)\\
\ge & \int_s E(Y\supzero \mid S\supzero=s, W=w)dF_1(s\mid w)-\int_s  E(Y\supzero \mid S\supzero=s, W=w)dF\supzero(s\mid w)\\
=& \int_s E(Y\supzero \mid S\supzero=s, W=w)d\left\{F\supone(s\mid w)-F\supzero(s\mid w)\right\}\\
=& \int_s \left\{F\supzero(s\mid w)-F\supone(s\mid w)\right\}\frac{\partial E(Y\supzero \mid S\supzero=s, W=w)}{\partial s}ds\ge 0,
\end{align*}
where $F^{(g)}(s\mid w)=P(S^{(g)}\le s|W=w), g=0, 1.$ That is, while treatment effect heterogeneity is allowed, the directions of the conditional average treatment effect among subgroups of patients with $W=w$ need to be consistent. One important implication is that under the null $H_0: \Delta=E\left\{\Delta(W)\right\}=0$, i.e., no average treatment effect, the conditional average treatment effect $\Delta(w)=0$ for all $w$ as well. Furthermore, from the derivation, it is clear that $\Delta(w)=0$ if and only if both
\begin{enumerate}
\item  $F\supone(s\mid w)=F\supzero(s\mid w)$, i.e., $P(S^{(1)} >s | W=w) = P(S^{(0)}  >s | W=w)$ and
\item  $E(Y^{(1)} | S^{(1)}=s,W=w) = E(Y^{(0)} | S^{(0)}=s,W=w).$
\end{enumerate}
Specifically, $\Delta(w)=0$ implies that there is no treatment effect on the distribution of the surrogate marker in the subgroup of patients with $W=w$.  In summary, under Assumptions (C1)-(C3)
$$\Delta=0 \Rightarrow \Delta(w)=0 \Rightarrow S\supone \mid W=w \sim S\supzero \mid W=w.$$
This relationship allows us to test the common null $H_0: \Delta=0$ via testing a seemingly more restrictive null that $S\supone \mid W=w \sim S\supzero \mid W=w,$ for all $w$ in the support of $W.$

For $(C2)$ and $(C3),$ if the primary outcome or surrogate are such that lower values are ``better", one can simply define the outcome/surrogate as $-X$ where $X$ is the initial value.

Assumptions $(C5)-(C6)$ are not required for the validity of the testing procedure proposed in the next section in that the p-value under the null follows a uniform distribution even without them, but it allows us to estimate a lower bound of the average treatment effect, $\Delta$, and construct the corresponding test statistic.

Under the following additional assumptions:
\begin{enumerate}
\item[] (C7) $Y \supone \perp S \supzero | S \supone, W$ and $Y \supzero \perp S \supone | S \supzero, W$;
\item[] (C8) $Y \suponep \perp S \supzerop | S \suponep, W^p$ and $Y \supzerop \perp S \suponep | S \supzerop, W^p,$
\end{enumerate}
 the treatment effect on the surrogate marker defined in Section \ref{test-statistic} and on the primary outcome can be interpreted within a causal framework: the proposed test statistic is an estimate of the portion of the treatment effect on the primary outcome attributable to the treatment effect on the surrogate marker. Otherwise, the proposed treatment effect on the surrogate marker can always serve as a lower bound for the average treatment effect on $Y$ and can be used in practice without assuming them.

To summarize,  Assumptions $(C1)-(C4)$ are needed for the validity of the proposed testing procedure, Assumptions $(C5)-(C6)$ allow us to interpret the test statistic based on he surrogate marker and baseline covariate only as a ``conservative'' estimator (or a lower bound) of the average treatment effect on the primary outcome, and causal interpretation of the lower is possible under additional assumptions $(C7)-(C8).$

\section*{Appendix C}
To estimate $\Delta$ using the primary outcome (gold standard) we use $\widehat{\Delta} = n_1^{-1} \sum_{i=1}^{n_1} Y_{1i} - n_0^{-1} \sum_{i=1}^{n_0} Y_{0i}$ and conduct a t-test to test $H_0: \Delta = 0$.

To estimate $\widetilde{\Delta}_P$, we use the nonparametric estimation approach of \cite{parast2019using} by estimating $\mu^p_0(s)$ as $$\widehat{\mu}^p_0(s) = \frac{\sum_{i=1}^{n_0^p} K_{h_4}(S^p_{0i} - s)Y^p_{0i}}{\sum_{i=1}^{n_0^p} K_{h_4}(S^p_{0i} - s)},$$ and then estimate $\widetilde{\Delta}_P$ as $$\widehat{\Delta}_P = n_1^{-1} \sum_{i=1}^{n_1} \widehat{\mu}^p_0(S_{1i}) -  n_0^{-1} \sum_{i=1}^{n_0} \widehat{\mu}^p_0(S_{0i}).$$ Note that this estimate only uses $S$ data from the current study (no $Y$ data from the current study) and $S,Y$ data from the previous study in group $Z=0$ only.  To obtain an estimate for the standard error of $\widehat{\Delta}_P$, $\sigma_P$, we simply take the empirical standard deviation of the transformed surrogate  i.e., let $\widetilde{Y}_{gi} =  \widehat{\mu}^p_0(S_{gi})$, and then $\widehat{\sigma}_P =  \widehat{var}(\widetilde{Y}_{1i})/n_1 + \widehat{var}(\widetilde{Y}_{0i})/n_0$ where $\widehat{var}$ indicates the empirical variance. This alternative testing procedure would then use the test statistic $Z_{P} = \widehat{\Delta}_{P}/\widehat{\sigma}_{P}$ and reject the null hypothesis when $|Z_P| > \Phi^{-1}(1-\alpha/2)$.

Importantly, one may also consider simply using the surrogate markers measured in the current study and define $\Delta_M = E(S\supone) - E(S\supzero)$ and conduct a t-test of $H_{0M}: \Delta_M=0$. The disadvantage of this approach is that there is no way to relate $\Delta_M$ and $\Delta$ i.e., the estimate of $\Delta_M$ does not give any helpful information about the magnitude of $\Delta$. In addition, this approach does not take advantage of information from the previous study nor does it account for heterogeneity in the utility of the surrogate marker. For these reasons, we do not compare our approach to this test.

\section*{Appendix D}



Our proposed estimator for $\widetilde{\Delta}_H$ is
$$ \widehat{\Delta}_{H} = \frac{1}{n}\left \{ \sum_{i=1}^{n_0} \left[\widehat{m}_1(W_{0i}; \widehat{\mu}_0^p)-\widehat{m}_0(W_{0i}; \widehat{\mu}_0^p)\right] + \sum_{i=1}^{n_1} \left[\widehat{m}_1(W_{1i}; \widehat{\mu}_0^p)-\widehat{m}_0(W_{1i}; \widehat{\mu}_0^p)\right] \right \}.$$
Let $\widetilde{\mu}_g=E\left\{\widehat{\mu}_0^{p}(S^{(g)}, W) \mid \widehat{\mu}_0^p \right\}, g=0, 1. $  It is obvious that $\widetilde{\Delta}_H=\widetilde{\mu}_1-\widetilde{\mu}_0.$  Also, let $m_g(w; \widehat{\mu}_0^p)=E\left\{\widehat{\mu}_0^p(S^{(g)}, W)\mid W=w\right\}.$

In this section, we only consider the randomness in the current study, i.e., the probability measure is conditional on $\widehat{\mu}_0^p(\cdot, \cdot).$ Now consider the centered term
\begin{align*}
 &\frac{1}{n}\sum_{g=0}^1\sum_{j=1}^{n_g} \widehat{m}_1(W_{gj}; \widehat{\mu}_0^p)-\widetilde{\mu}_1\\
=&\frac{1}{n}\sum_{g=0}^1 \sum_{j=1}^{n_g}\left[n_1^{-1}\sum_{i=1}^{n_1} \frac{K_h(W_{1i}-W_{gj})\widetilde{S}_{1i}}{\widehat{f}_1(W_{gj})}\right]-\widetilde{\mu}_1,
 \end{align*}
which is
\begin{align*}
 &\frac{1}{nn_1} \sum_{j=1}^{n_0}\sum_{i=1}^{n_1} \frac{K_h(W_{1i}-W_{0j})\widetilde{S}_{1i}}{\widehat{f}_1(W_{0j})}+ \frac{1}{n}\sum_{i=1}^{n_1} \left[ \frac{1}{n_1}\sum_{j=1}^{n_1} \frac{K_h(W_{1i}-W_{1j})}{\widehat{f}_1(W_{1j})}\right]\widetilde{S}_{1i}-\widetilde{\mu}_1 \\
 =&\frac{1}{nn_1} \sum_{j=1}^{n_0}\sum_{i=1}^{n_1} \frac{K_h(W_{1i}-W_{0j})\widetilde{S}_{1i}}{\widehat{f}_1(W_{0j})}+ \frac{1}{n}\sum_{i=1}^{n_1} \left[ \frac{1}{n_1}\sum_{j=1}^{n_1} K_h(W_{1i}-W_{1j})\right]\frac{\widetilde{S}_{1i}}{\widehat{f}_1(W_{1i})}-\widetilde{\mu}_1+O_p(h^2) \\
 =&\frac{n_0}{nn_1}\sum_{i=1}^{n_1} \frac{\widehat{f}_0(W_{1i})}{\widehat{f}_1(W_{1i})} \widetilde{S}_{1i}+ \frac{1}{n}\sum_{i=1}^{n_1}\widetilde{S}_{1i}-\widetilde{\mu}_1+O_p(h^2)\\
 =&\frac{1}{n_1}\sum_{i=1}^{n_1}(\widetilde{S}_{1i}-\widetilde{\mu}_1)+ \frac{n_0}{nn_1}\sum_{i=1}^{n_1} \frac{\widehat{f}_0(W_{1i})-\widehat{f}_1(W_{1i})}{\widehat{f}_1(W_{1i})} \widetilde{S}_{1i}+O_p(h^2)\\
 =&\frac{1}{n_1}\sum_{i=1}^{n_1}(\widetilde{S}_{1i}-\widetilde{\mu}_1)+ \frac{n_0}{nn_1}\sum_{i=1}^{n_1} \left[\frac{1}{n_0}\sum_{j=1}^{n_0} K_h(W_{0j}-W_{1i})-\frac{1}{n_1}\sum_{j=1}^{n_1}K_h(W_{1j}-W_{1i})\right]\frac{\widetilde{S}_{1i}}{f_1(W_{1i})}+O_p(h^2)\\
 =&\frac{1}{n_1}\sum_{i=1}^{n_1}(\widetilde{S}_{1i}-\widetilde{\mu}_1)+ \pi_0 \left[\frac{1}{n_0}\sum_{i=1}^{n_0} \widehat{m}_1(W_{0i}; \widehat{\mu}_0^p)-\frac{1}{n_1}\sum_{i=1}^{n_1}\widehat{m}_1(W_{1i}; \widehat{\mu}_0^p)\right]+O_p(h^2)\\
  =&\frac{1}{n_1}\sum_{i=1}^{n_1}(\widetilde{S}_{1i}-\widetilde{\mu}_1)+ \pi_0 \left[\frac{1}{n_0}\sum_{i=1}^{n_0} m_1(W_{0i}; \widehat{\mu}_0^p)-\frac{1}{n_1}\sum_{i=1}^{n_1}m_1(W_{1i}; \widehat{\mu}_0^p)\right]\\
  &+\pi_0\left[\frac{1}{n_0}\sum_{i=1}^{n_0}\left(\widehat{m}_1(W_{0i}; \widehat{\mu}_0^p)- m_1(W_{0i}; \widehat{\mu}_0^p)\right)-\frac{1}{n_1}\sum_{i=1}^{n_1}\left(\widehat{m}_1(W_{1i}; \widehat{\mu}_0^p)-m_1(W_{1i}; \widehat{\mu}_0^p)\right)\right]+O_p(h^2)
 \end{align*}
where $\pi_g=n_g/n$ and $\widehat{f}_1(w)$ is the nonparametric estimator for the density function of $W$ based on observations in treatment group 1. Now, consider the expansion
$$ \widehat{m}_1(w; \widehat{\mu}_0^p)-m_1(w; \widehat{\mu}_0^p)=\frac{1}{n_1}\sum_{i=1}^{n_1} K_h(W_{1i}-w)\left\{\widetilde{S}_{1i}-m_1(W_{1i}; \widehat{\mu}_0^p)\right\}+O_p\left(h^2+\frac{\log(n_1)}{n_1h}\right)$$
uniform in $w.$ Therefore,
\begin{align*}
&\frac{1}{n_0}\sum_{j=1}^{n_0}\left\{\widehat{m}_1(W_{0j}; \widehat{\mu}_0^p)- m_1(W_{0j}; \widehat{\mu}_0^p)\right\}\\
=& \frac{1}{n_1n_0}\sum_{j=1}^{n_0}\sum_{i=1}^{n_1} K_h(W_{1i}-W_{0j})\left\{\widetilde{S}_{1i}-m_1(W_{1i}; \widehat{\mu}_0^p)\right\}+O_p\left(h^2+\frac{\log(n_1)}{n_1h}\right)\\
=& \frac{1}{n_1}\sum_{i=1}^{n_0}\widehat{f}_0(W_{1i})\left\{\widetilde{S}_{1i}-m_1(W_{1i}; \widehat{\mu}_0^p)\right\}+O_p\left(h^2+\frac{\log(n_1)}{n_1h}\right)\\
=& \frac{1}{n_1}\sum_{i=1}^{n_0}f_0(W_{1i})\left\{\widetilde{S}_{1i}-m_1(W_{1i}; \widehat{\mu}_0^p)\right\}+O_p\left(h^2+\frac{\log(n_1)}{n_1h}\right)+o_p\left(\frac{1}{\sqrt{n_1}}\right)
\end{align*}
Similarly,
\begin{align*}
&\frac{1}{n_1}\sum_{i=1}^{n_1}\left(\widehat{m}_1(W_{1i}; \widehat{\mu}_0^p)-m_1(W_{1i}; \widehat{\mu}_0^p)\right)\\
=&\frac{1}{n_1}\sum_{i=1}^{n_0}f_0(W_{1i})\left\{\widetilde{S}_{1i}-m_1(W_{1i}; \widehat{\mu}_0^p)\right\}+O_p\left(h^2+\frac{\log(n_1)}{n_1h}\right)+o_p\left(\frac{1}{\sqrt{n_0}}\right),
\end{align*}
and
\begin{align}
  &\sqrt{n}\left[\frac{1}{n_0}\sum_{i=1}^{n_0}\left(\widehat{m}_1(W_{0i}; \widehat{\mu}_0^p)- m_1(W_{0i}; \widehat{\mu}_0^p)\right)-\frac{1}{n_1}\sum_{i=1}^{n_1}\left(\widehat{m}_1(W_{1i}; \widehat{\mu}_0^p)-m_1(W_{1i}; \widehat{\mu}_0^p)\right)\right]\\
 =&O_p\left(\sqrt{n_1}h^2+\frac{\log(n_1)}{\sqrt{n_1}h}\right)+o_p(1). \label{residual}
 \end{align}
Therefore, when $h=O(n_1^{-\delta}), \delta\in (1/4, 1/2)$, the right hand side of (\ref{residual}) becomes $o_p(1),$ and thus
\begin{align*}
&\frac{1}{\sqrt{n}}\sum_{g=0}^1\sum_{j=1}^{n_g} \widehat{m}_1(W_{gj}; \widehat{\mu}_0^p)-\widetilde{\mu}_1\\
=&\frac{\sqrt{n}}{n_1}\sum_{i=1}^{n_1}(\widetilde{S}_{1i}-\widetilde{\mu}_1)+ \pi_0 \left[\frac{\sqrt{n}}{n_0}\sum_{j=1}^{n_0} m_1(W_{0j}; \widehat{\mu}_0^p)-\frac{\sqrt{n}}{n_1}\sum_{j=1}^{n_1}m_1(W_{1j}; \widehat{\mu}_0^p)\right]+o_p(1).
\end{align*}
Finally, we have
\begin{align*}
&\sqrt{n}\left\{\widehat{\Delta}_H-\widetilde{\Delta}_H\right\}\\
=&\frac{\sqrt{n}}{n_1}\sum_{i=1}^{n_1}(\widetilde{S}_{1i}-\widetilde{\mu}_1)+ \pi_0 \left[\frac{\sqrt{n}}{n_0}\sum_{i=1}^{n_0} m_1(W_{0i}; \widehat{\mu}_0^p)-\frac{\sqrt{n}}{n_1}\sum_{i=1}^{n_1}m_1(W_{1i}; \widehat{\mu}_0^p)\right]\\
&-\frac{\sqrt{n}}{n_0}\sum_{i=1}^{n_0}(\widetilde{S}_{0i}-\widetilde{\mu}_0)+ \pi_1 \left[\frac{\sqrt{n}}{n_1}\sum_{i=1}^{n_1} m_0(W_{1i}; \widehat{\mu}_0^p)-\frac{\sqrt{n}}{n_0}\sum_{i=1}^{n_0}m_0(W_{0i}; \widehat{\mu}_0^p)\right]+o_p(1)\\
=&\frac{\sqrt{n}}{n_1}\sum_{i=1}^{n_1}\left(\widetilde{S}_{1i}-\pi_0m_1(W_{1i}; \widehat{\mu}_0^p)-\pi_1m_0(W_{1i}; \widehat{\mu}_0^p)-\pi_1(\widetilde{\mu}_1-\widetilde{\mu}_0)\right)\\
 &-\frac{\sqrt{n}}{n_0}\sum_{i=1}^{n_0}\left(\widetilde{S}_{0i}- \pi_0 m_1(W_{0i}; \widehat{\mu}_0^p)-\pi_1 m_0(W_{0i}; \widehat{\mu}_0^p)-\pi_0(\widetilde{\mu}_1-\widetilde{\mu}_0)\right)+o_p(1),
\end{align*}
which converges weakly to a mean zero Gaussian distribution with a variance of
$$ \frac{1}{\pi_1} E\left\{\widetilde{S}_{1i}-\pi_0m_1(W_{1i}; \widehat{\mu}_0^p)-\pi_1m_0(W_{1i}; \widehat{\mu}_0^p)-\pi_1\widetilde{\Delta}_H\right\}^2$$
$$+\frac{1}{\pi_0} E\left\{\widetilde{S}_{0i}- \pi_0 m_1(W_{0i}; \widehat{\mu}_0^p)-\pi_1 m_0(W_{0i}; \widehat{\mu}_0^p)-\pi_0\widetilde{\Delta}_H\right\}^2.$$
Therefore, the variance of $\widehat{\Delta}_H$ can be estimated as
$$\widehat{\sigma}^2_H=\frac{1}{n_1^2}\sum_{i=1}^{n_1}\left(\widetilde{S}_{1i}- \pi_0 \widehat{m}_1(W_{1i}; \widehat{\mu}_0^p))-\pi_1\widehat{m}_0(W_{1i}; \widehat{\mu}_0^p)-\pi_1\widehat{\Delta}_H)\right)^2$$
$$+\frac{1}{n_0^2}\sum_{i=1}^{n_0}\left(\widetilde{S}_{0i}- \pi_0 \widehat{m}_1(W_{0i}; \widehat{\mu}_0^p)-\pi_1 \widehat{m}_0(W_{0i}; \widehat{\mu}_0^p)-\pi_0\widehat{\Delta}_H\right)^2$$

Next, we will derive the asymptotical distribution of  $\sqrt{n}(\widehat{\Delta}_H^{AUG}-\widetilde{\Delta}_H)$. It is clear that
\begin{align*}
&\sqrt{n}(\widehat{\Delta}_H^{AUG}-\widetilde{\Delta}_H)\\
=&\frac{\sqrt{n}}{n_1}\sum_{i=1}^{n_1} \left\{\widetilde{S}_{1i}-\pi_0 \widehat{m}_1(W_{1i}; \widehat{\mu}_0^p)-\pi_1 \widehat{m}_0(W_{1i}; \widehat{\mu}_0^p)-\pi_1\widetilde{\Delta}_H\right\}\\
 &-\frac{\sqrt{n}}{n_0}\sum_{i=1}^{n_1} \left\{\widetilde{S}_{0i}-\pi_0 \widehat{m}_1(W_{0i}; \widehat{\mu}_0^p)-\pi_1 \widehat{m}_0(W_{0i}; \widehat{\mu}_0^p)-\pi_0\widetilde{\Delta}_H\right\}\\
=&\frac{\sqrt{n}}{n_1}\sum_{i=1}^{n_1} \left\{\widetilde{S}_{1i}-\pi_0 m_1(W_{1i}; \widehat{\mu}_0^p)-\pi_1 m_0(W_{1i}; \widehat{\mu}_0^p)-\pi_1\widetilde{\Delta}_H\right\}\\
 &-\frac{\sqrt{n}}{n_0}\sum_{i=1}^{n_1} \left\{\widetilde{S}_{0i}-\pi_0 m_1(W_{0i}; \widehat{\mu}_0^p)-\pi_1 m_0(W_{0i}; \widehat{\mu}_0^p)-\pi_0\widetilde{\Delta}_H\right\}\\
&-\sqrt{n}\left[\frac{\pi_0}{n_0}\sum_{i=1}^{n_0}\left(\widehat{m}_1(W_{0i}; \widehat{\mu}_0^p)- m_1(W_{0i}; \widehat{\mu}_0^p)\right)-\frac{\pi_0}{n_1}\sum_{i=1}^{n_1}\left(\widehat{m}_1(W_{1i}; \widehat{\mu}_0^p)-m_1(W_{1i}; \widehat{\mu}_0^p)\right)\right]\\
&-\sqrt{n}\left[\frac{\pi_1}{n_1}\sum_{i=1}^{n_1}\left(\widehat{m}_1(W_{1i}; \widehat{\mu}_0^p)- m_1(W_{1i}; \widehat{\mu}_0^p)\right)-\frac{\pi_1}{n_0}\sum_{i=1}^{n_0}\left(\widehat{m}_0(W_{0i}; \widehat{\mu}_0^p)-m_1(W_{0i}; \widehat{\mu}_0^p)\right)\right]\\
=&\frac{\sqrt{n}}{n_1}\sum_{i=1}^{n_1} \left\{\widetilde{S}_{1i}-\pi_0 m_1(W_{1i}; \widehat{\mu}_0^p)-\pi_1 m_0(W_{1i}; \widehat{\mu}_0^p)-\pi_1\widetilde{\Delta}_H\right\}\\
 &-\frac{\sqrt{n}}{n_0}\sum_{i=1}^{n_1} \left\{\widetilde{S}_{0i}-\pi_0 m_1(W_{0i}; \widehat{\mu}_0^p)-\pi_1 m_0(W_{0i}; \widehat{\mu}_0^p)-\pi_0\widetilde{\Delta}_H\right\}+o_p(1)\\
=& \sqrt{n}(\widehat{\Delta}_H-\widetilde{\Delta}_H)+o_p(1).
\end{align*}
Therefore, $\widehat{\Delta}_H^{AUG}$ and $\widehat{\Delta}_H$ are asymptotically equivalent.
Furthermore, noting that
\begin{align*}
&\widetilde{S}_{1i}-\pi_0 m_1(W_{1i}; \widehat{\mu}_0^p)-\pi_1 m_0(W_{1i}; \widehat{\mu}_0^p)-\pi_1\widetilde{\Delta}_H\\
=&\left\{\widetilde{S}_{1i}-m_1(W_{1i}; \widehat{\mu}_0^p)\right\}+\pi_1\left\{m_1(W_{1i}; \widehat{\mu}_0^p)- m_0(W_{1i}; \widehat{\mu}_0^p)-\widetilde{\Delta}_H\right\}
\end{align*}
and
$$E\left[\left\{\widetilde{S}_{1i}-m_1(W_{1i}; \widehat{\mu}_0^p)\right\}\left\{m_1(W_{1i}; \widehat{\mu}_0^p)- m_0(W_{1i}; \widehat{\mu}_0^p)-\widetilde{\Delta}_H\right\} \mid W_{1i} \right]=0,$$
we have
\begin{align*}
&E\left[\widetilde{S}_{1i}-\pi_0 m_1(W_{1i}; \widehat{\mu}_0^p)-\pi_1 m_0(W_{1i}; \widehat{\mu}_0^p)-\pi_1\widetilde{\Delta}_H\right]^2\\
=& E\left[\widetilde{S}_{1i}- m_1(W_{1i}; \widehat{\mu}_0^p)\right]^2+ \pi_1^2 E\left[m_1(W_{1i}; \widehat{\mu}_0^p)-m_0(W_{1i}; \widehat{\mu}_0^p)-\widetilde{\Delta}_H\right]^2.
\end{align*}
Similarly,
\begin{align*}
&E\left[\widetilde{S}_{0i}-\pi_0 m_1(W_{0i}; \widehat{\mu}_0^p)-\pi_1 m_0(W_{0i}; \widehat{\mu}_0^p)-\pi_0\widetilde{\Delta}_H\right]^2\\
=&E\left[\widetilde{S}_{0i}- m_0(W_{0i}; \widehat{\mu}_0^p)\right]^2+ \pi_0^2 E\left[ m_1(W_{0i}; \widehat{\mu}_0^p)-m_0(W_{0i}; \widehat{\mu}_0^p)-\widetilde{\Delta}_H\right]^2.
\end{align*}
Therefore, the variance of $\widehat{\Delta}_H^{(AUG)}$ can also be consistently estimated by
$$\widehat{\sigma}_{AUG}^2=\frac{1}{n_1^2}\sum_{i=1}^{n_1} \left[\widehat{\mu}_{0}^{(p)}(S_{1i}, W_{1i})- \widehat{m}_1(W_{1i}; \widehat{\mu}_0^p)\right]^2 +\frac{1}{n_0^2}\sum_{i=1}^{n_0} \left[\widehat{\mu}_{0}^{(p)}(S_{0i}, W_{0i})-\widehat{m}_0(W_{0i}; \widehat{\mu}_0^p)\right]^2$$
$$+\frac{\pi_1^2}{n_1^2}\sum_{i=1}^{n_1}\left[\widehat{m}_1(W_{1i}; \widehat{\mu}_0^p)- \widehat{m}_0(W_{1i}; \widehat{\mu}_0^p)-\widehat{\Delta}_H\right]^2+\frac{\pi_0^2}{n_0^2}\sum_{i=1}^{n_0}\left[\widehat{m}_1(W_{0i}; \widehat{\mu}_0^p)- \widehat{m}_0(W_{0i}; \widehat{\mu}_0^p)-\widehat{\Delta}_H\right]^2,$$
and $\widehat{\Delta}_{(AUG)}/\widehat{\Delta}_H=1+o_p(1).$

\section*{Appendix E}
Here, we provide an example where there is heterogeneity in the utility of the surrogate and the $W$ is distributed the \textit{same} in the prior study and current study, but $\Delta_P$ still fails to provide a lower bound for $\Delta$. In both the prior study and the current study, we assume that $\log(W)\sim \epsilon_W$,  $S^{(g)}=W\times \exp(\delta_0g+\epsilon_S)$, and $Y^{(g)}=S^{(g)} W, g\in \{0, 1\},$ where $\delta_0$ is a positive constant, and $\epsilon_W$ and $\epsilon_S$ are two independent standard normals. It is obvious that
$\mu_0^p(s, w)=sw$
and
\begin{align*}
\Delta=\Delta_H=&E(S\supone W)-E(S\supzero W)=E\left\{W E(S\supone-S\supzero\mid W)\right\}\\
=&E\left\{W\left(\exp(0.5+\delta_0)W-\exp(0.5)W \right)\right\}=\exp\left(\frac{5}{2}\right)\left(\exp(\delta_0)-1\right).
\end{align*}
Next, we have
\begin{align*}
\mu_0^p(s)=& E(W S\supzero \mid  S\supzero =s)= s E(W\supzero \mid S\supzero=s)\\
=& s\times \exp\left(\frac{1}{4}\right)s^{\frac{1}{2}}= \exp\left(\frac{1}{4}\right)s^{\frac{3}{2}},
\end{align*}
and
\begin{align*}
\Delta_P=& E\left\{\left(S\supone\right)^{\frac{3}{2}}\exp\left(\frac{1}{4}\right)\right\}-E\left\{\left(S\supzero\right)^{\frac{3}{2}}\exp\left(\frac{1}{4}\right)\right\}\\
=&\exp\left(\frac{5}{2}\right)\left(\frac{3\delta_0}{2}-1 \right).
\end{align*}
Consequently, in this setting,  $\Delta_P>\Delta=\Delta_H$ even though the $W$ has the same distribution in both studies.

\clearpage

\end{document}